\PassOptionsToPackage{table}{xcolor}
\documentclass{opendatalab}

\usepackage[utf8]{inputenc}
\usepackage[T1]{fontenc}
\usepackage{amsmath}
\usepackage{array}
\usepackage{float}
\usepackage{titletoc}
\usepackage{enumitem}
\usepackage{tabularx}
\usepackage{longtable}
\usepackage{listings}
\usepackage{makecell}
\usepackage{xspace}
\usepackage{colortbl}
\usepackage{pifont}

\newcommand{\github}{\raisebox{-1.5pt}{\includegraphics[height=1.05em]{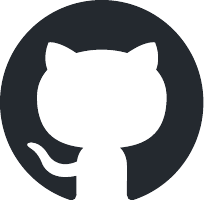}}\xspace}
\newcommand{\web}{\raisebox{-1.5pt}{\includegraphics[height=1.05em]{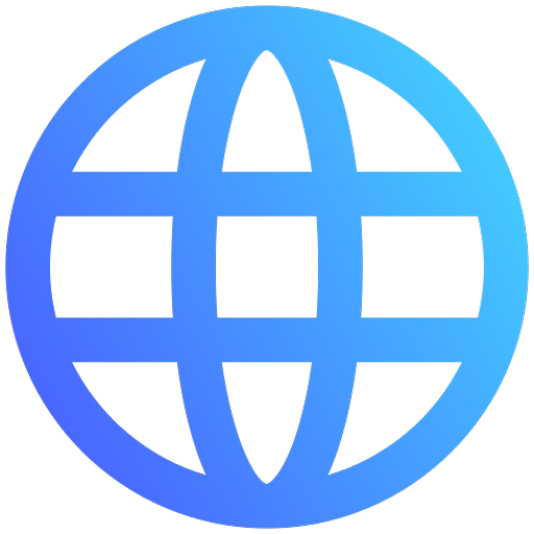}}\xspace}
\newcommand{\huggingface}{\raisebox{-1.5pt}{\includegraphics[height=1.05em]{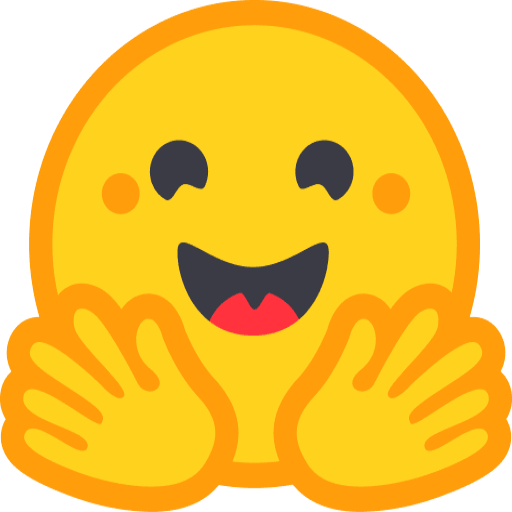}}\xspace}

\definecolor{promptcolor}{HTML}{EAF3FF}
\definecolor{promptcolorheader}{HTML}{8DB7E8}
\definecolor{codeback}{HTML}{F7FAFF}

\newtcolorbox{promptbox}[1][]{
  enhanced,
  breakable,
  top=0.3em,
  bottom=0.3em,
  left=0.5em,
  right=0.5em,
  toptitle=0.3em,
  bottomtitle=0.2em,
  boxsep=0pt,
  colframe=promptcolorheader,
  colback=promptcolor!50,
  boxrule=0.5pt,
  width=\columnwidth,
  title={\footnotesize #1}
}

\title{PaperFlow: Profiling, Recommending, and Adapting Across Daily Paper Streams}

\author[1]{Fuqiang Wang}
\author[1]{Song Tan}
\author[1]{Zheng Guo}
\author[1]{Jiaohao Fu}
\author[2]{Xinglong Xu}
\author[2]{Bihui Yu}
\author[2]{Jie Dong}
\author[2]{Zheng Sun}
\author[1]{Siyuan Li}
\author[1]{Jingxuan Wei}
\author[3]{Cheng Tan}

\affiliation[1]{Key Laboratory of Computing Power Network and Information Security, Ministry of Education, Shandong Computer Science Center (National Supercomputer Center in Jinan), Qilu University of Technology (Shandong Academy of Sciences)}
\affiliation[2]{University of Chinese Academy of Science}
\affiliation[3]{Shanghai Artificial Intelligence Laboratory}

\abstract{
Scientific paper recommendation is typically evaluated as static ranking over a fixed candidate set, yet real scientific reading unfolds as a daily, longitudinal process in which interests shift and feedback accumulates. We introduce PaperFlow, a framework that organizes it into three coupled stages: \textit{Profiling}, which constructs and maintains a structured, inspectable scholarly profile from heterogeneous cold-start evidence; \textit{Recommending}, which ranks each date-specific paper stream through multi-signal aggregation under a fixed display budget; and \textit{Adapting}, which updates user state from semantically distinct feedback signals and models interest drift across days. We further define a longitudinal user-day benchmark that fixes users, dates, candidate pools, visible inputs, and hidden simulated relevance labels under a shared temporal information boundary. The benchmark contains 24 simulated research users, 50 daily paper streams, 1,200 user-day episodes, 20,727 unique papers, and 497,448 episode–paper records. We additionally specify a blind human-evaluation protocol to validate alignment between automatic metrics and expert judgments. Experiments against five scientific recommendation baselines show that PaperFlow achieves the strongest oracle-based ranking, the highest behavioral alignment with simulated reading selections, and the best blind human-evaluation score.
}

\metadatainfo{
    \small\makebox[\linewidth][c]{
        \github~\href{https://github.com/OpenRaiser/PaperFlow}{\textbf{Code}} \quad
        \web~\href{https://openraiser.github.io/PaperFlow}{\textbf{Website}} \quad
        \huggingface~\href{https://huggingface.co/datasets/OpenRaiser/PaperFlow}{\textbf{Dataset}}
    }
}

\begin{document}

\maketitle

\section{Introduction}

\begin{figure*}[t]
\centering
\includegraphics[width=\textwidth]{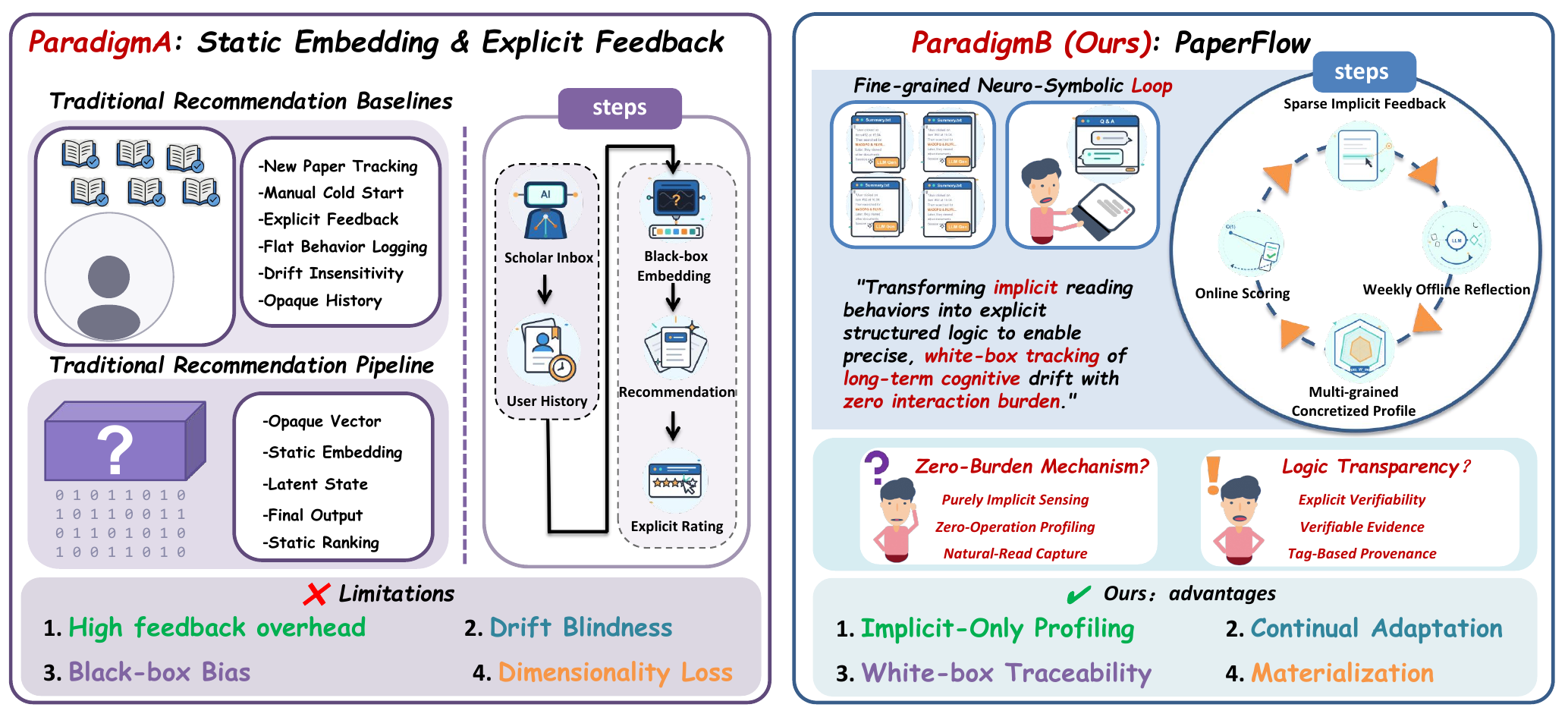}
\caption{Motivation and paradigm comparison between traditional scientific paper recommendation and PaperFlow.}
\label{fig:motivation}
\end{figure*}

Scientific paper recommendation is typically framed as a one-shot ranking problem: given a user representation and a fixed candidate set, produce a relevance-sorted list. This formulation overlooks the temporal structure of real scientific reading. In practice, researchers confront a fresh paper stream every day, allocate scarce reading time to a small subset, skip many plausible papers, and continuously revise their interests as projects mature and new methods emerge~\cite{wei2026trinity,yu2026paperfit,pan2026programming,wei2026pager}. A system that serves this workflow must solve three tightly coupled problems: it must \textit{profile} the researcher, maintaining a structured, inspectable representation of evolving interests; it must \textit{recommend}, selecting from today's candidate stream under a strict display budget; and it must \textit{adapt}, updating user state from semantically heterogeneous feedback so that each day's reading informs the next day's ranking. Figure~\ref{fig:motivation} illustrates this shift from static recommendation to a daily scientific reading loop.

Existing work addresses parts of this loop, but usually in separate settings. Scholar Inbox is the closest predecessor, studying daily scientific recommendations with explicit feedback and cold-start initialization \citep{flicke2025scholar}; natural-language profile and paper-assistant systems improve profile transparency, search, question answering, survey generation, and reading support \citep{ramos2024transparent,wang2024surveyagent,fok2023scim}; and dynamic recommendation work models feedback and interest drift in evolving user representations \citep{lin2025interest}. However, these efforts do not yet define a controlled temporal user-day task: one in which methods rank the same daily candidate pools using only visible pre-ranking inputs, consume cross-day feedback and drift state in strict temporal order, and are evaluated against hidden relevance labels without future information.

We propose PaperFlow, a framework that couples the three stages within a single daily loop. For \textit{profiling}, PaperFlow maintains a structured, inspectable scholarly profile that separates research-interest state, controllable preferences, and drift state into distinct, editable fields. For \textit{recommending}, it ranks each date-specific candidate pool through multi-signal aggregation, combining semantic matching, author and institution priors, behavioral signals, and explicit rules under a fixed display budget. For \textit{adapting}, it updates user state with signal-specific semantics: selections, explicit edits, and sustained reading provide strong interest evidence, whereas skips supply only weak, context-dependent signals; a drift module uses cross-day behavior to distinguish transient exploration from sustained migration. In parallel, a reading-report channel provides post-selection reading assistance whose feedback adjusts report organization and evidence density rather than directly changing research-interest weights.

We further construct a longitudinal benchmark for user-day evaluation. Each episode fixes a simulated research user, a date, a candidate pool, visible pre-ranking inputs, and historical feedback under a shared temporal boundary. Methods rank the same frozen candidate pools; same-day selections, post-recommendation states, drift outcomes, and hidden simulated relevance labels remain unavailable until evaluation, preventing future-information leakage. The labels enable reproducible comparison but are not human-annotated ground truth. The current snapshot contains 24 simulated users, 50 days, 1,200 user-day episodes, 20,727 unique papers, and 497,448 episode--paper records. A blind human-evaluation protocol is defined separately to validate alignment between automatic metrics and expert judgments. Together, this work contributes a sequential user-day formulation of scientific paper recommendation, the PaperFlow framework that couples structured profiling, multi-signal daily ranking, signal-aware state updating, and behavior-driven drift modeling within a single loop. Experiments against five external baselines that characterize main-setting gains and the trade-off between static relevance ranking and dynamic adaptation.

\section{Related Work}

\subsection{Scientific Paper Recommendation}

Scientific paper recommendation must jointly address relevance estimation, cold start, and profile interpretability. Scholar Inbox is the workflow-level predecessor, combining daily paper recommendations with explicit feedback and cold-start initialization~\citep{flicke2025scholar}; but it relies on opaque embeddings and does not model how interests evolve after initialization. Content-based methods improve paper matching through user-behavior modeling~\citep{xu2025information}, while OMRC-MR strengthens scientific-paper representations through QA-style discourse summarization, multi-level contrastive learning, and structure-aware re-ranking~\citep{wang2025discourse}. Citation-enhanced and entity-enhanced methods incorporate external impact signals or fine-grained scientific knowledge~\citep{liu2025citation,liu2024kucnet}. Contextualized paper alerts and natural-language profiles increase transparency and support user-specific explanation~\citep{lee2024paperweaver, ramos2024transparent}. Recommender-agent and conversational systems further explore tool use, multi-turn decision making, and preference elicitation~\citep{huang2024recommenderagent, zhao2024toollearning,wang2024recmind,zhang2024agentcf, shu2024rah,gao2023chatrec,fang2024macrs,li2024chatcrs}. PaperFlow differs from these approaches by maintaining a structured, updateable scholarly profile across daily recommendation and feedback loops rather than treating profile evidence primarily as one-time initialization input.

\subsection{Paper Reading Assistants}

Paper reading assistants support paper search, survey generation, synthesis, and reading comprehension. SurveyAgent and PaSa organize retrieval, recommendation, query generation, and paper screening for academic search \citep{wang2024surveyagent,he2025pasa}. OpenScholar, Arxiv Copilot, and scientific language agents support retrieval-augmented synthesis and personalized academic assistance \citep{asai2024openscholar,lin2024papercopilot,skarlinski2024languageagents}. Reading-support systems cover interactive scholarly reading, intelligent skimming, localized citation contexts, and mixed-initiative synthesis \citep{lo2023semantic,fok2023scim,rachatasumrit2022citeread,kang2023synergi}. Socially grounded systems bring paper recommendation into research group contexts \citep{wang2025paperping,wang2025socialrag}. PaperFlow instead treats cross-day reading as the primary unit, organizing selections, skips, explicit corrections, intensive-reading requests, report feedback, and profile updates into user-day episodes. 

\subsection{Dynamic Feedback and Interest Drift}

Dynamic feedback and interest-drift studies emphasize that personalization must distinguish short-term exploration from stable preference migration. IDURL models interest drift in sequential recommendation \citep{lin2025interest}, while PISA studies the stability-plasticity trade-off in continual recommendation \citep{yoo2025pisa}. Planning and feedback-loop frameworks formulate recommendation as multi-round interaction optimization \citep{shi2024planners,cai2024flow}. Agentic recommender benchmarks emphasize cold start, evolving interests, dynamic information acquisition, and preference updating \citep{shang2025agentrecbench,huang2026recbenchplus,liang2026pdrbench}. PaperFlow grounds these issues in scientific reading, where feedback signals have different semantics and adaptation must be evaluated under date-frozen paper streams.
\par

\section{PaperFlow Method}

\subsection{System Overview}

\begin{figure*}[t]
\centering
\makebox[\textwidth][c]{%
\includegraphics[width=1.06\textwidth,trim=0 6 20 6,clip]{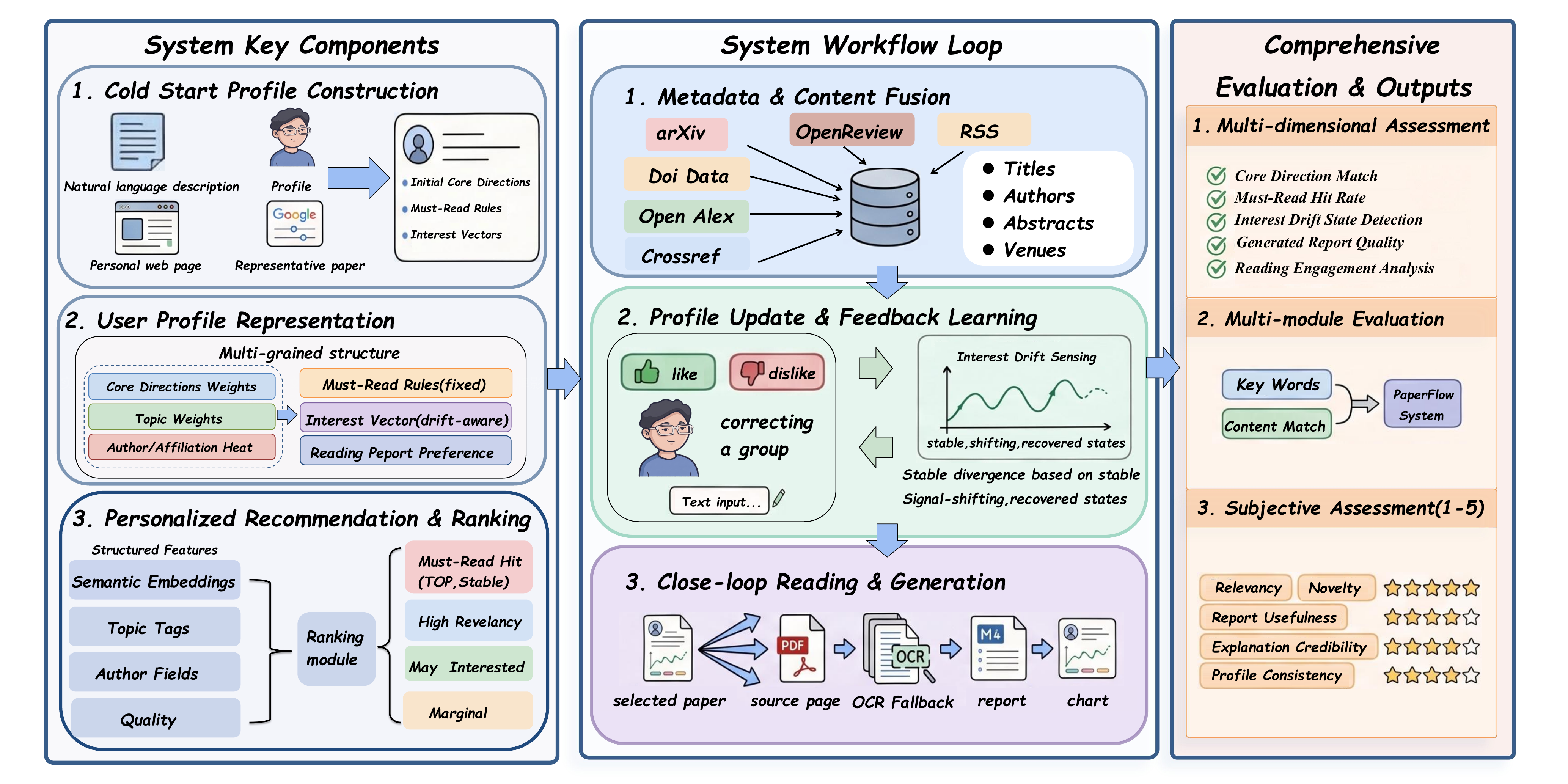}
}
\caption{Overview of the PaperFlow dynamic personalized scientific reading loop.}
\label{fig:method-framework}
\end{figure*}

PaperFlow models paper recommendation as a dynamic personalized reading loop. As shown in Figure~\ref{fig:method-framework}, it constructs a structured scholarly profile from cold-start evidence, ranks each daily candidate pool, and folds selections, skips, reading requests, report feedback, and profile edits back into the next-day profile and drift state, with a reading-report channel for post-selection assistance.

\subsection{Structured Scholarly Profile}

Let $t$ index a daily recommendation round, i.e., day $t$. PaperFlow maintains an editable and continuously updated structured scholarly profile $p_t$ at the beginning of day $t$, decomposed as follows:
\begin{equation}
p_t = \{D_t, T_t, A_t, I_t, M_t, v_t, \tau_t, Q_t, B_t, d_t\},
\end{equation}
where $D_t$, $T_t$, $v_t$ capture research directions, topic weights, and semantic interest vectors; $A_t$, $I_t$, $M_t$ encode author, institution, and must-read priors; $\tau_t$ stores method and paper-type preferences; and $Q_t$, $B_t$, $d_t$ represent report preferences, reading-behavior signals, and the drift-adaptation state. We separate $Q_t$ from $B_t$ so that report-style feedback does not contaminate research-interest weights.

When no historical feedback exists, PaperFlow builds $p_0$ from multi-source cold-start evidence, including research descriptions, profile pages, representative papers or PDFs, and manual preferences. The benchmark uses reproducible simulated profiles, while these sources describe broader cold-start inputs supported by the method.
\par

Cold-start construction extracts and canonicalizes directions, methods, application contexts, authors, institutions, and preference clues from heterogeneous sources. The LLM supports extraction and canonicalization; repeated support raises initial weight or confidence, explicit inputs remain inspectable rules or fields, and all outputs must fit fixed profile fields and pass structural validation before entering shared state.

\subsection{Daily Updating}

On day $t$, PaperFlow ranks the date-specific candidate pool $C_t$ with the current profile $p_t$, aggregating long-term interest matching, prior signals, dynamic behavioral signals, and explicit rules.

For a candidate paper $c$, the ranking signal is decomposed as:
\begin{equation}
\begin{aligned}
\operatorname{score}(c,p_t) ={}&
S_{\text{match}}(c,p_t) + S_{\text{prior}}(c,p_t) \\
&+ S_{\text{dyn}}(c,p_t) + S_{\text{must}}(c,p_t).
\end{aligned}
\end{equation}
The terms group matching, priors, dynamics, and must-read rules: $S_{\text{match}}$ covers semantic-interest similarity and topic matching; $S_{\text{prior}}$ covers author, institution, and candidate-paper quality; $S_{\text{dyn}}$ covers drift state, anchor directions, old-topic suppression, and recent reading behavior; and $S_{\text{must}}$ covers must-read rules. The feature weights and normalization are in Appendix~\ref{app:ranking-score-parameters}. Candidate-paper quality in $S_{\text{prior}}$ is distinct from profile-local $Q_t$.

Recommendations use four system-side display and diagnostic tiers: \texttt{must\_read}, \texttt{high\_relevant}, \texttt{maybe\_interested}, and \texttt{edge\_relevant}. Non-must-read tiers use score and rank-aware thresholds; must-read matches receive a small bonus while preserving personalized relevance.
\par

Feedback updating keeps signal semantics separate: selections, corrections, and profile edits provide strong interest evidence; skips provide weak negative evidence; reading requests, PDF uploads, and repeated reading update $B_t$; and report feedback updates only $Q_t$. Research-interest updates mainly use selections, sustained reading, explicit edits, and cross-day drift evidence.
\par

We abstract feedback-driven updating as:
\begin{equation}
p_{t+1} = U(p_t, F_t, B_t, d_t),
\end{equation}
where $p_{t+1}$ is the next-round profile, $U$ is the update process, $F_t$ is daily feedback, $B_t$ is the reading-behavior state, and $d_t$ is the adaptation strength between long- and short-term evidence.

\subsection{Behavior-Driven Interest Drift}

PaperFlow uses behavioral evidence to distinguish transient exploration from sustained interest migration. Repeated selections, explicit requests, intensive reading, PDF uploads, and manual profile edits can support a new direction, while unsupported old directions are gradually decayed in the profile and ranking process.
\par

Let $L_t$ be the long-term topic distribution, $S_t$ the recent short-term topic distribution, $F_t$ the feedback set for the day, and $B_t$ the reading-behavior state. PaperFlow computes an internal drift-evidence signal:
\begin{equation}
g_t = \operatorname{Drift}(L_t,S_t,F_t,B_t),
\end{equation}
which measures the deviation between long-term interest and recent behavior and controls drift-state transitions and profile-update strength. It increases under sustained evidence for a low-weight new direction and decreases when evidence weakens or realigns with the long-term profile. To prevent a single interaction from causing profile jumps, PaperFlow bounds the per-round update of each topic direction $z$:
\begin{equation}
\left|T_{t+1}(z) - T_t(z)\right| \leq \epsilon,
\end{equation}
where $\epsilon$ is the per-round upper bound on topic-weight change.

PaperFlow uses four drift states. \textit{Stable} means recent behavior matches the long-term profile; \textit{Observing} marks an emerging direction with insufficient evidence; \textit{Shifting} increases sustained new-topic weight under the update constraint; and \textit{Recovered} rebalances new and old interests after confirmed migration. Transitions are not triggered by a single click; they depend on drift-evidence thresholds, sustained evidence windows, and directional consistency. Detailed threshold and window settings are given in Appendix~\ref{app:interest_drift}.

\begin{figure*}[t]
\centering
\vspace{-2mm}
\makebox[\textwidth][c]{%
\includegraphics[width=1.02\textwidth,trim=8 4 8 4,clip]{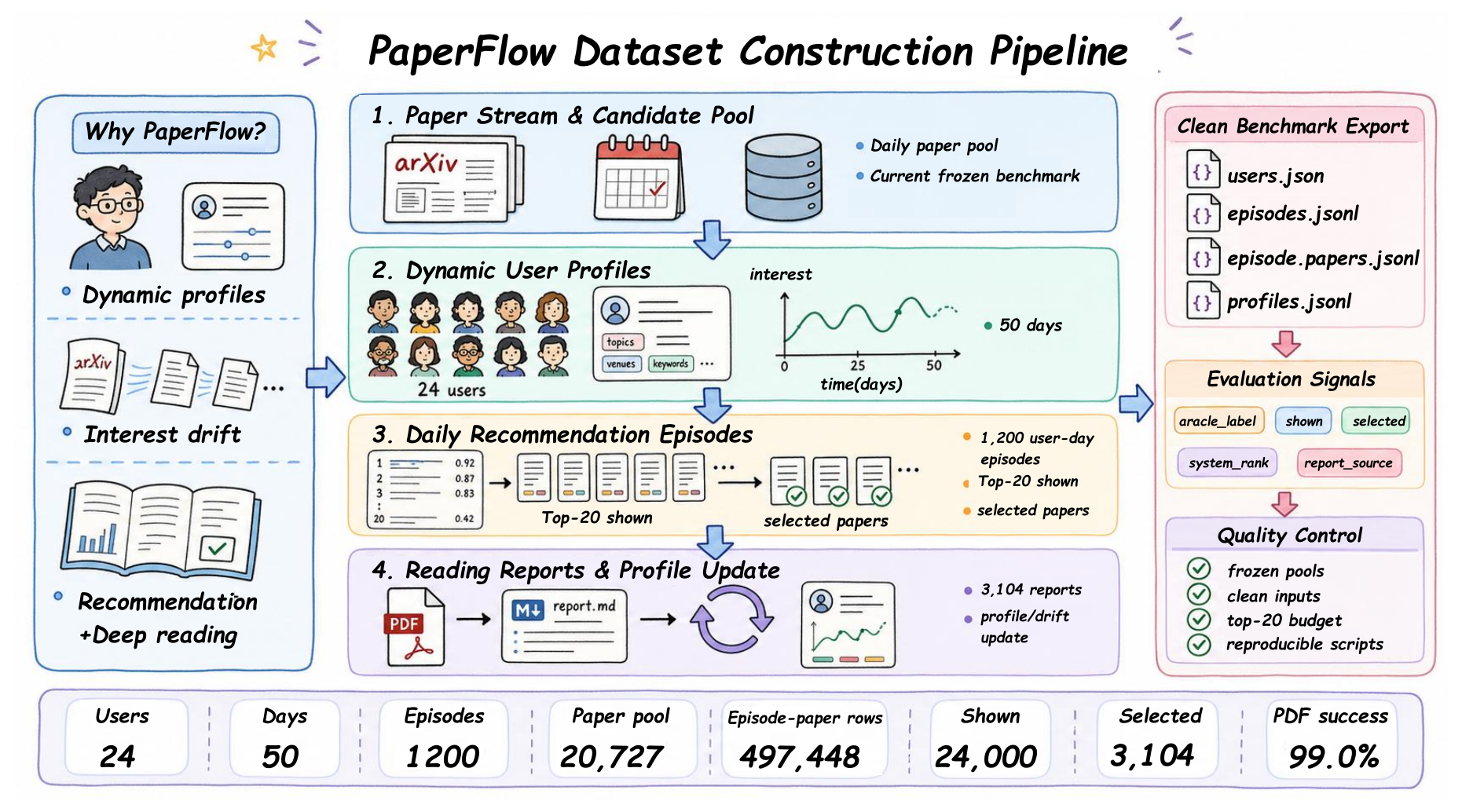}
}
\vspace{-6mm}
\caption{Construction pipeline of the PaperFlow benchmark. Daily paper streams and simulated researcher profiles are converted into date-frozen user-day episodes, providing clean method inputs alongside hidden pseudo-oracle labels, reading-report records, and drift diagnostics.}
\label{fig:benchmark-construction}
\end{figure*}

\section{PaperFlow Benchmark}
\label{sec:benchmark}

\subsection{Benchmark Construction}

As shown in Figure~\ref{fig:benchmark-construction}, PaperFlow is constructed as a longitudinal reading pipeline rather than a static set of user--paper pairs. The pipeline freezes daily paper pools, initializes dynamic researcher profiles, executes daily recommendation episodes, records reading and profile-update traces, and exports clean benchmark files with evaluation signals kept separate from method inputs.

\noindent\textbf{Paper stream and candidate pool.}
PaperFlow converts the arXiv daily paper stream into date-indexed candidate pools. Each date-specific pool contains only papers visible that day and is frozen before evaluation, strictly isolating ranking from temporal data leakage and eliminating variation from dynamic crawling or later metadata updates.

\noindent\textbf{Dynamic user profiles.}
To establish reproducible cold-start conditions, the pipeline initializes 24 simulated researcher profiles spanning scientific domains. Each profile contains structured interests, preferences, and drift plans, so episodes can test both stable personalization and controlled interest migration without relying on private user logs.

\noindent\textbf{Daily recommendation episodes.}
The benchmark then rolls profiles through the frozen paper stream to produce user-day episodes. For each episode, a method observes the allowed pre-ranking state and returns a Top-20 list from $C_t$; the simulator records shown papers, selected papers, hidden relevance labels, system ranks, and reading signals for later analysis and state updates.

\noindent\textbf{Reading reports, profile updates, and export.}
Selected papers trigger reading-report generation and profile/drift updates, producing the downstream traces needed to study continuous use. The final export separates clean inputs from oracle labels, shown/selected flags, report sources, and quality-control fields, enforcing a strict temporal boundary for comparable evaluation. Appendix~\ref{app:benchmark} gives the complete field specification.

\subsection{Benchmark Analysis}
Figure~\ref{fig:benchmark-construction} summarizes the scale of the frozen snapshot. PaperFlow contains 24 simulated researchers, 50 daily paper pools, and 1,200 user-day episodes. The 20,727-paper pool expands to 497,448 episode--paper records, creating a daily filtering problem in which hundreds of candidates must be compressed into a Top-20 reading list.
A Top-20 list shows only about 4.8\% of an average daily pool, and the simulator selects 3,104 of the 24,000 shown papers for deeper reading. This creates sparse but temporally ordered feedback: most papers are merely skipped, while selected papers generate reading reports and profile-update evidence. 



\begin{table*}[t]
  \centering
  \caption{Main results on the PaperFlow benchmark.
  Automatic metrics are computed on the full 1,200 user-day episodes with a
  Top-20 display budget. All scores are reported on a 0--100 scale; ratio
metrics are multiplied by 100. HumanEval is a blind listwise human score on
  sampled Top-20 lists.}
  \vspace{-2mm}
  \label{tab:main-results}
  \small
  \renewcommand{\arraystretch}{1.1}
  \begin{tabularx}{\textwidth}{@{} l *{6}{>{\centering\arraybackslash}X} @{}}
  \toprule
  \textbf{Method}
    & \makecell{\textbf{gNDCG}\\@20}
    & \makecell{\textbf{Useful}\\@5}
    & \makecell{\textbf{Useful}\\@20}
    & \makecell{\textbf{Selected}\\NDCG@20}
    & \makecell{\textbf{Rec.}\\Score}
    & \textbf{HumanEval} \\
\midrule
  Scholar Inbox
    & 39.00 & 25.92 & 14.63 & 33.47 & 46.30 & 55.56 \\
  Citation-Enhanced
    & 32.67 & 24.45 & 12.68 & 34.24 & 42.34 & 44.44 \\
  OMRC-MR
    & 19.95 & 15.52 & 8.68 & 25.59 & 27.37 & 30.00 \\
  UPR
    & 37.54 & 26.43 & 13.85 & 29.74 & 43.70 & 53.33 \\
  KUCNet
    & 33.14 & 23.93 & 12.69 & 33.47 & 42.27 & 35.56 \\
  \midrule
  \rowcolor{cyan!12}
  \textbf{PaperFlow}
    & \textbf{50.65} & \textbf{34.90} & \textbf{17.56}
    & \textbf{70.88} & \textbf{55.31} & \textbf{65.56} \\
  \bottomrule
  \end{tabularx}
  \vspace{-4mm}
\end{table*}

\subsection{Evaluation Protocol and Labels}

PaperFlow evaluates personalized scientific reading as strictly sequential user-day ranking. For each simulated user and date, a method produces:
\begin{equation}
\begin{aligned}
R_t &= \operatorname{Alg}(u,t,x_u,C_t,F_{<t},q_{u,<t}), \\
&\text{s.t.}\quad |R_t| = 20,\quad R_t \subseteq C_t,
\end{aligned}
\end{equation}
where $x_u$ is visible user metadata, $C_t$ is the date-frozen candidate pool, $F_{<t}$ contains only earlier feedback, and $q_{u,<t}$ is the pre-ranking dynamic state. Same-day selections, oracle labels, and drift outcomes remain hidden until evaluation.

All methods rank the same users, dates, and candidate pools under a fixed-round offline protocol. Automatic evaluation uses controlled pseudo-oracle relevance labels (\texttt{strong\_relevant}, \texttt{relevant}, \texttt{weak\_relevant}, \texttt{irrelevant}); these labels provide reproducible evaluation targets but are not human ground truth.

We report three complementary metric families: (1)~oracle-based ranking metrics (e.g., gNDCG@20, OracleRecall@20, MRR@20), summarized by a composite \texttt{RecommendationScore} ranging from 0 to 100; (2)~\texttt{SelectedNDCG@20}, measuring agreement with simulated reading selections; and (3)~\texttt{HumanEval}, a blind listwise score over sampled anonymized Top-20 lists to validate automatic--human alignment. All metrics are computed strictly after recommendation and never exposed as same-day inputs. Detailed definitions and annotator agreement are in Appendices~\ref{app:evaluation-metrics} and~\ref{sec:human-eval-protocol}.

\section{Experiments}
\label{sec:experiments}





\subsection{Baselines}
\label{sec:main-baselines}

We compare PaperFlow with five representative scientific-paper
recommendation baselines: Scholar Inbox \citep{flicke2025scholar},
Citation-Enhanced Literature \citep{liu2025citation},
OMRC-MR Content \citep{wang2025discourse},
Natural-Language User Profile \citep{ramos2024transparent},
and KUCNet Enhanced Recommendation \citep{liu2024kucnet}.
All methods rank the same candidate pools under the same Top-20
budget. The compared methods differ in whether they support cold-start profiling,
profile updating, interest drift, feedback use, and reading-report assistance, summarized in
Appendix~\ref{app:capability-coverage}.

\subsection{Main Recommendation Results}
\label{sec:main-results}

Table~\ref{tab:main-results} shows that PaperFlow achieves the strongest
automatic performance and the highest blind listwise human score. Compared
with Scholar Inbox, the strongest external baseline by
\texttt{RecommendationScore}, PaperFlow improves \texttt{gNDCG@20} from
39.00 to 50.65, \texttt{RecommendationScore} from 46.30 to 55.31, and
\texttt{HumanEval} from 55.56 to 65.56.

The largest qualitative difference appears in behavior alignment:
\texttt{SelectedNDCG@20} increases from 33.47 to 70.88. This suggests that
PaperFlow not only retrieves relevant papers under hidden labels, but also
ranks papers closer to those the simulated user later chooses for reading.
Other baselines recover useful signals in specific settings, but remain
weaker on overall ranking quality and downstream behavioral alignment.

\subsection{LLM Comparison Results}
\label{sec:llm-backbone-comparison}

We compare LLM backbones within PaperFlow, using Gemini 3 Flash as the default. The frozen benchmark, user profiles, candidate pools, Top-20 budget, embedding model, and metric computation are fixed; only the LLM used for structured judgment, recommendation explanation, feedback parsing, and reading-report generation is changed. We report quality, human alignment, and cost separately: \texttt{ModelAutoScore} combines recommendation and report quality, \texttt{ModelHumanScore} provides the corresponding blind human score. Full definitions are given in Appendix~\ref{sec:appendix-report-auto-score}, ~\ref{sec:appendix-model-auto-score}, and ~\ref{sec:human-eval-protocol}.

\begin{table}[ht]
  \centering
  \caption{LLM backbone comparison under PaperFlow.}
  \label{tab:llm-comparison}
  \small
  \setlength{\tabcolsep}{1.6pt}
  \begin{tabular*}{\columnwidth}{@{\extracolsep{\fill}}
    p{0.10\columnwidth}
    p{0.34\columnwidth}
    rrrr
    @{}}
  \toprule
  \textbf{Group} & \textbf{Model} &
  \makecell{\textbf{Rec.}\\Score} &
  \makecell{\textbf{Report}\\Auto} &
  \makecell{\textbf{Model}\\Auto} &
  \makecell{\textbf{Model}\\Human} \\
  \midrule
  \multirow{8}{*}{\makecell[c]{Closed}}
  & GPT-5.4 & 54.77 & 99.31 & 63.68 & 69.41 \\
  & Qwen3.5-Plus & 55.23 & 99.18 & 64.02 & 74.07 \\
  & Gemini 3.1 Pro  & 55.09 & 99.10 & 63.89 & 70.30 \\
  & Claude Sonnet 4.6 & 55.36 & 99.28 & 64.15 & 82.07 \\
  & Qwen3.6-Plus & 55.22 & 99.80 & 64.14 & 77.11 \\
  & Qwen3.6-Max & 55.11 & 99.21 & 63.93 & 74.15 \\
  & Grok 4.3 & 56.31 & 99.56 & 64.96 & 94.07 \\
  \cmidrule(l){2-6}
  & \cellcolor{cyan!12}Gemini 3 Flash
    & \cellcolor{cyan!12}{55.31}
    & \cellcolor{cyan!12}{99.70}
    & \cellcolor{cyan!12}{64.19}
    & \cellcolor{cyan!12}{76.74} \\
  \midrule
  \multirow{6}{*}{\makecell[c]{Open}}
  & MiMo-V2.5-Pro & 54.89 & 99.94 & 63.90 & 73.93 \\
  & DeepSeek-V4-Pro & 55.16 & 99.52 & 64.04 & 76.44 \\
  & DeepSeek-V4-Flash & 55.43 & 99.95 & 64.34 & 82.15 \\
  & Kimi K2.6 & 55.91 & 99.82 & 64.69 & 86.67 \\
  & GLM-5.1 & 55.03 & 99.77 & 63.98 & 75.41 \\
  & MiniMax-M2.7 & 55.48 & 99.19 & 64.22 & 81.78 \\
  \bottomrule
  \end{tabular*}
  \vspace{-4mm}
\end{table}

\begin{figure}[ht]
  \centering

  \begin{minipage}[t]{0.48\columnwidth}
    \centering
    \includegraphics[width=\linewidth,trim=40 24 0 8,clip]{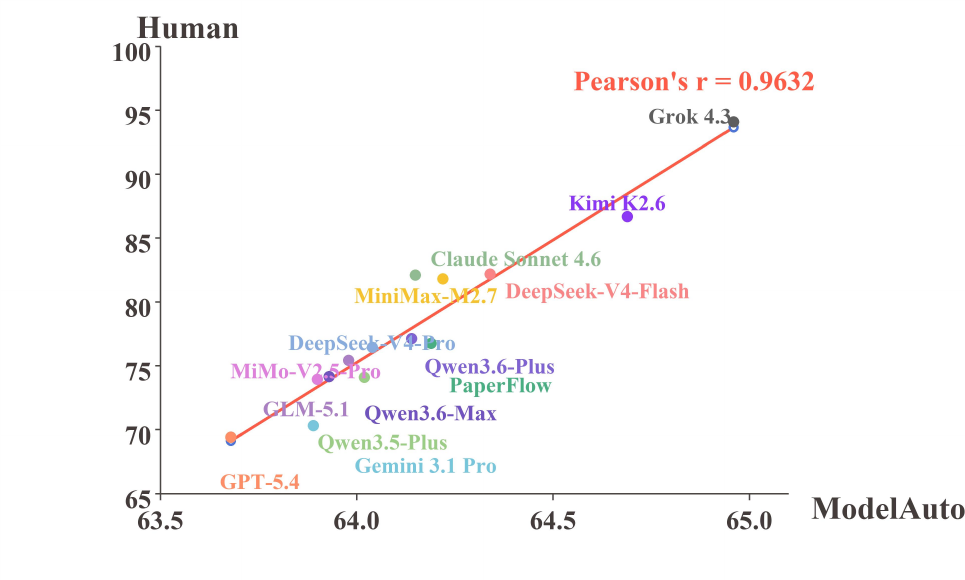}
    \caption{Automatic--human metric alignment (\texttt{ModelAutoScore} vs. \texttt{ModelHumanScore}).}
    \label{fig:model-auto-human-alignment}
  \end{minipage}
  \hfill
  \begin{minipage}[t]{0.48\columnwidth}
    \centering
    \includegraphics[width=\linewidth,trim=0 8 0 5,clip]{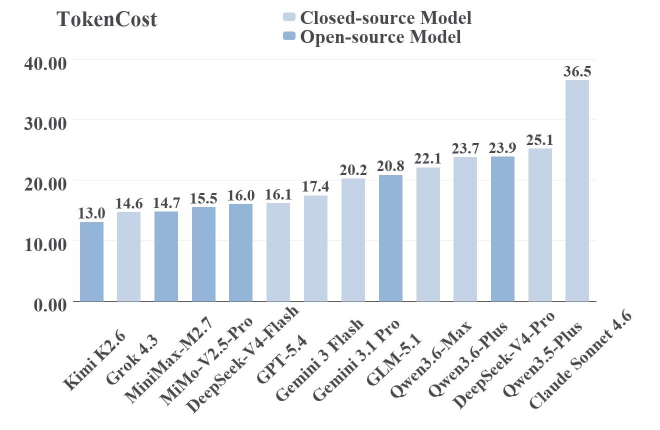}
    \caption{Token-cost across LLM backbones. Bar values are reported in million tokens; lower is better.}
    \label{fig:llm-token-cost}
  \end{minipage}

  \vspace{-2mm}
\end{figure}

Table~\ref{tab:llm-comparison} shows that model choice materially affects both automatic and human-aligned quality under identical retrieval, embedding, and evaluation conditions. Among closed-source models, Grok~4.3 achieves the highest \texttt{RecommendationScore} (56.31) and \texttt{ModelHumanScore} (94.07) while also maintaining the lowest token cost (Figure~\ref{fig:llm-token-cost}), making it the most cost-effective closed-source option. Gemini~3~Flash offers the best overall balance as default backbone with strong \texttt{ModelAutoScore} (64.19) and competitive human alignment. 

Figure~\ref{fig:model-auto-human-alignment} plots \texttt{ModelAutoScore} against \texttt{ModelHumanScore} for all backbones, yielding Pearson's $r{=}0.9632$. This strong correlation validates automatic metrics as a reliable proxy for human judgment and suggests that improvements observed in automatic evaluation
translate consistently to perceived recommendation quality. Figure~\ref{fig:llm-token-cost} reports token cost in millions across all LLM backbones. The results suggest that token efficiency is largely model-dependent rather than correlated with quality.

\begin{table*}[t]
  \centering
  \caption{Ablation and diagnostic results. Simplified variants can improve
oracle-based relevance concentration, while \texttt{SelectedNDCG@20} measures
alignment with simulated downstream reading selections. Full PaperFlow performs
best on this behavior-alignment metric.}
  \vspace{-2mm}
  \label{tab:ablation-results}
  \small
  \renewcommand{\arraystretch}{1.15}
  \setlength{\tabcolsep}{4.5pt}
  \begin{tabular}{@{} l c c c c c c c c c @{}}
  \toprule
  \textbf{Method}
    & \makecell{\textbf{gNDCG}\\@20\,$\uparrow$}
    & \makecell{\textbf{Useful}\\@5\,$\uparrow$}
    & \makecell{\textbf{Useful}\\@20\,$\uparrow$}
    & \makecell{\textbf{Oracle}\\Rec@20\,$\uparrow$}
    & \makecell{\textbf{Lift}\\@20\,$\uparrow$}
    & \makecell{\textbf{StrictR}\\@20+\,$\uparrow$}
    & \makecell{\textbf{MRR}\\@20\,$\uparrow$}
    & \makecell{\textbf{Selected}\\NDCG@20\,$\uparrow$}
    & \makecell{\textbf{Rec.}\\Score\,$\uparrow$} \\
  \midrule
  \rowcolor{cyan!12}
  \textbf{PaperFlow}
    & 0.5065 & 0.3490 & 0.1756 & 0.2232 & 12.73 & 0.8613
    & 0.6044 & \textbf{0.7088} & 55.31 \\
  \midrule
  Fixed Profile
    & \textbf{0.5346} & \textbf{0.3710} & \textbf{0.1877}
    & \textbf{0.2302} & \textbf{13.36} & \textbf{0.8881}
    & 0.6265 & 0.6955 & \textbf{57.81} \\
  w/o Explicit Pref.
    & 0.5148 & 0.3643 & 0.1731
    & 0.1971 & 12.86 & 0.7607
    & \textbf{0.6435} & 0.6659 & 54.37 \\
  w/o Drift
    & 0.5147 & 0.3543 & 0.1804
    & 0.2298 & 12.95 & 0.8869
    & 0.6066 & 0.7039 & 56.36 \\
  w/o Reading Signal
    & 0.5126 & 0.3537 & 0.1786
    & 0.2278 & 12.90 & 0.8789
    & 0.6053 & 0.7074 & 56.06 \\
  \bottomrule
  \end{tabular}
  \vspace{-2mm}
\end{table*}

\begin{figure*}[t]
  \centering
  \includegraphics[width=\textwidth,trim=4 8 4 8,clip]{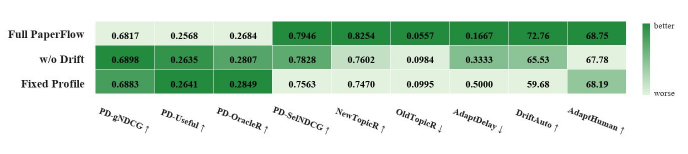}
  \caption{Interest-drift analysis. Cell color is normalized within each
  metric, with darker green indicating better performance. PostDrift metrics
  are computed on the post-drift window; NewTopicR, OldTopicR, AdaptDelay,
  DriftAuto, and AdaptHuman summarize adaptation-oriented behavior.}
  \label{fig:interest-drift-results}
  \vspace{-2mm}
\end{figure*}

\subsection{Ablation Analysis}
\label{sec:ablation-analysis}

Table~\ref{tab:ablation-results} separates oracle-based relevance from
behavioral alignment. Several simplified variants obtain higher oracle-based
scores than Full PaperFlow, because removing profile updates, drift modeling,
or reading signals makes the system more conservative and closer to stable
pseudo-oracle labels.

In contrast, Full PaperFlow achieves the best \texttt{SelectedNDCG@20}, showing
stronger agreement with simulated downstream reading selections. The ablation
results therefore reveal a static--dynamic trade-off: simplified variants can
improve static relevance concentration, while PaperFlow's adaptive components
better track what users choose to read over time.

\subsection{Interest Drift Analysis}
\label{sec:interest-drift-analysis}

We evaluate PaperFlow on controlled interest-drift episodes, comparing PaperFlow with w/o Drift and Fixed Profile. The goal is to test whether the system exposes new-interest papers, suppresses stale old-topic exposure, adapts quickly, and remains aligned with later reading choices.

Figure~\ref{fig:interest-drift-results} shows that w/o Drift and Fixed Profile can slightly improve some PostDrift oracle metrics, but PaperFlow is strongest on adaptation-oriented signals: it achieves the highest PostDrift \texttt{SelectedNDCG@20}, the highest \texttt{NewTopicRecall@20}, the lowest \texttt{OldTopicRate@20}, and the shortest adaptation delay. PaperFlow also has the best \texttt{DriftAutoScore} (72.76) and \texttt{Adaptation\allowbreak Human\allowbreak Score} (68.75), indicating that the drift module improves adaptation rather than merely maximizing static oracle relevance. The drift-score and human-evaluation protocol are in Appendix~\ref{sec:appendix-drift-auto-score} and Appendix~\ref{sec:human-eval-protocol}.

\subsection{Real-User Pilot Study}
\label{sec:real-user-pilot}

We conduct a real-user pilot study with five graduate students in computer science. Each participant uses PaperFlow for 5--7 interaction rounds over daily paper pools. The study records actual reading decisions and post-round Likert ratings; participant information, metric definitions, questionnaire items, and per-user results are provided in Appendix~\ref{app:real-user-pilot}. As shown in Table~\ref{tab:real-user-pilot}, the real-user results mirror the simulated benchmark: PaperFlow consistently outperforms baselines in both actual reading-selection rates and user satisfaction.

\begin{table}[ht]
  \centering
  \vspace{-2mm} 
  \caption{Real-user results averaged over five participants. Precision@$k$
  and ReadRate are computed from actual reading decisions; Satisfaction is a
  1--5 Likert score.}
  \vspace{-1mm} 
  \label{tab:real-user-pilot}
  \small
  \setlength{\tabcolsep}{3pt}
  \renewcommand{\arraystretch}{1.05}
  \begin{tabular*}{\columnwidth}{@{\extracolsep{\fill}}lrrrr@{}}
  \toprule
  Method & Prec@5 & Prec@20 & ReadRate & Sat. \\
  \midrule
  Daily arXiv Email & 0.40 & 0.26 & 0.18 & 2.8 \\
  Static Profile & 0.58 & 0.38 & 0.27 & 3.5 \\
  PaperFlow & \textbf{0.71} & \textbf{0.47} & \textbf{0.34} & \textbf{4.0} \\
  \bottomrule
  \end{tabular*}
  \vspace{-3mm}
\end{table}

\subsection{Case Study}
\label{sec:case-study}

\begin{figure}[ht]
  \centering
  \includegraphics[width=0.8\columnwidth,trim=0 4 0 4,clip]{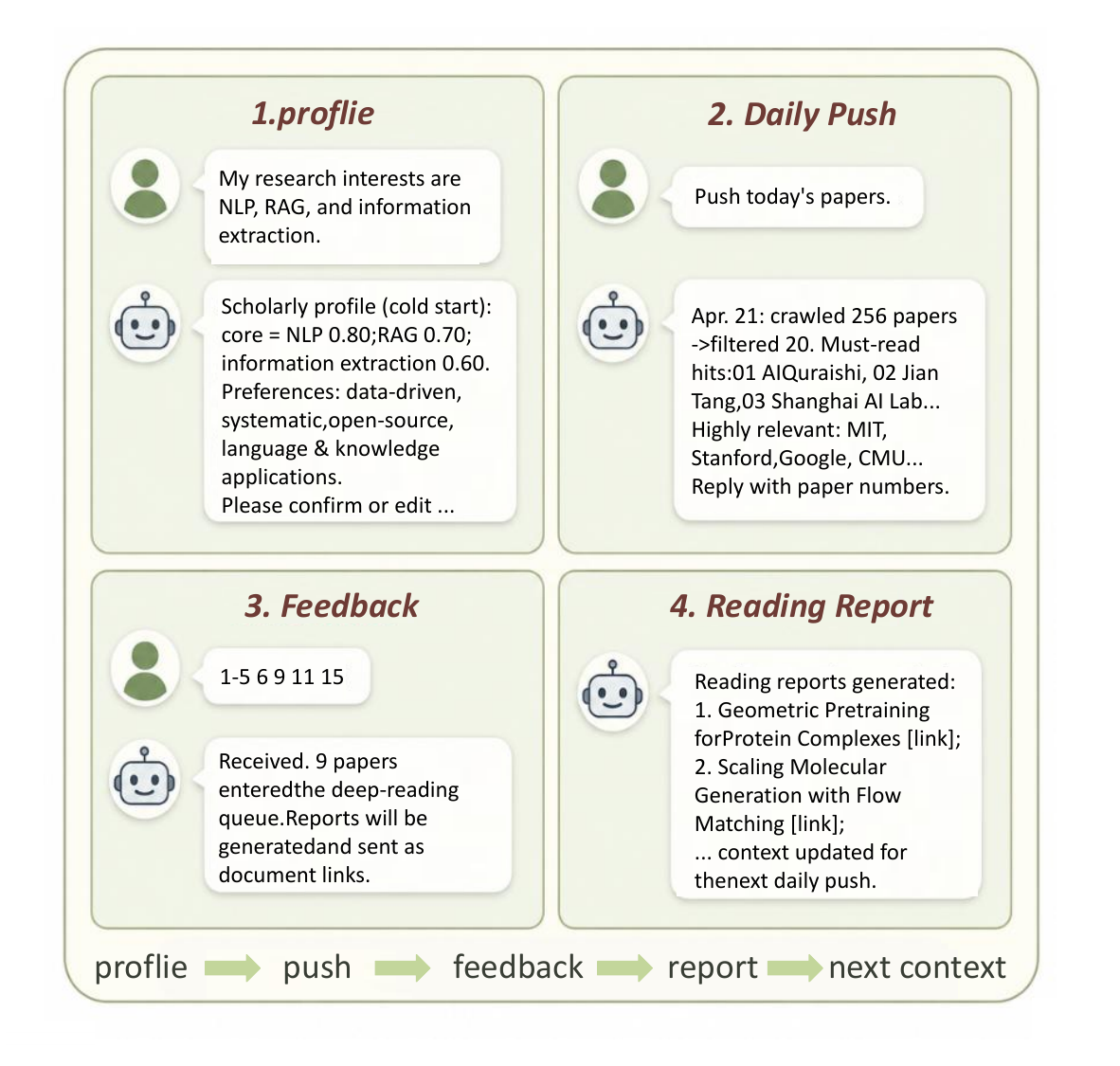}
  \vspace{-8mm}
  \caption{Representative PaperFlow case study. 
  }
  \label{fig:case-study}
\end{figure}

We present a representative interaction to illustrate how PaperFlow operates as
a closed-loop reading assistant rather than a one-shot ranker.
Figure~\ref{fig:case-study} shows four stages in one daily use case: cold-start
profile construction, daily paper push, user feedback, and reading-report
generation. The example demonstrates how user feedback is converted into
reading signals and then carried forward as context for the next recommendation
round.

PaperFlow converts research-interest descriptions into an editable scholarly profile. It then filters the daily paper stream into a compact candidate list, accepts feedback through paper numbers, and generates reading reports for the selected papers. 

\section{Conclusion}

We introduced PaperFlow for personalized paper reading, combining structured scholarly profiles, multi-signal daily ranking, feedback-driven state updates, and interest-drift modeling. The benchmark fixes longitudinal user-day episodes, candidate pools, hidden relevance labels, and diagnostic trajectories under a shared temporal boundary.
\section*{Limitations}

PaperFlow primarily uses simulated research users and simulated relevance
labels. Although we include a real-user pilot study, it is intended as a
user-experience sanity check. This design provides reproducibility and controlled temporal
comparison, but the simulated labels should not be interpreted as
human-annotated truth or as deployment logs from real users. The current
benchmark is mainly derived from arXiv daily paper streams, so
coverage may differ across fields and publication venues. Future work should
connect the protocol with larger-scale human evaluation, additional scholarly
sources, and deployment studies.



\bibliographystyle{plainnat}
\setcitestyle{numbers}
\bibliography{ref}

\clearpage
\beginappendix

\section*{Appendix Table of Contents}
\begingroup
\hypersetup{linkcolor=black}
\startcontents[appendix]
\printcontents[appendix]{}{1}{\setcounter{tocdepth}{2}}
\endgroup

\vspace{1em}

\clearpage




\section{Benchmark and Data Setting}
\label{app:benchmark}
\subsection{Benchmark Scope}

We use the same PaperFlow benchmark snapshot for the main experiment, baseline comparison, and LLM comparison. During evaluation, the paper pool is not expanded. Instead, all methods operate on a pre-built historical paper database. This design isolates the effect of the recommendation method itself and avoids incomparability caused by paper-pool expansion, retrieval-time differences, or data updates.

\begin{table*}[ht]
  \centering
  \caption{Benchmark setting used by PaperFlow.}
  \label{tab:PaperFlow-benchmark-setting}
  \small
  \setlength{\tabcolsep}{4pt}
  \renewcommand{\arraystretch}{1.08}
  \begin{tabular*}{\textwidth}{@{\extracolsep{\fill}}p{0.26\textwidth}p{0.68\textwidth}@{}}
  \toprule
  Item & Setting \\
  \midrule
  Data directory &
  \texttt{data/benchmark\_full\_24users\_20260301\_20260419}\\
  & \texttt{\_show20\_with\_reading} \\
  Date range & March 1, 2026 to April 19, 2026 \\
  Number of users & 24 simulated researchers \\
  Number of episodes & 1,200 user-day episodes \\
  Average candidate-pool size & 414.54 papers per episode \\
  Display budget & Top-20 \\
  Average number of selected papers & 2.59 papers per episode \\
  Paper-pool policy & Fixed paper pool; no new papers are added during evaluation \\
  Main-experiment LLM & \texttt{Gemini 3 Flash Preview} \\
  Embedding model & \texttt{Qwen3-Embedding-8B} \\
  Random seed & 42 \\
  Reading reports & Enabled for selected papers \\
  \bottomrule
  \end{tabular*}
  \vspace{-2mm}
\end{table*}

Each episode represents the recommendation process for one user on one day. The system first ranks the candidate papers using the user profile, paper content, must-read rules, interest-drift state, and recent reading signals. It then shows the Top-20 papers and simulates the user's selection of a small number of papers for subsequent reading. Each episode records the user-profile snapshot, candidate-pool statistics, Top-20 list, system labels, oracle labels, user selections, drift state, reading-signal state, and token usage.

This fixed-pool design has two advantages. First, it ensures that all methods are compared on exactly the same candidate sets, avoiding data bias introduced by dynamic crawling or real-time retrieval. Second, it preserves temporal order. The system must not only rank a single candidate pool, but also maintain user profiles, react to interest changes, and accumulate reading signals across consecutive days. The benchmark therefore evaluates both static ranking quality and longitudinal user modeling.

\subsection{Episode and Subsession Definition}

We treat each user-day episode as one subsession. A subsession is not a separate model prompt; it is the logging and contextual unit used by the experiment. It contains the full recommendation context for a user on a specific day, including the user profile, candidate pool, Top-20 list, user selections, interest-drift state, and reading reports.

\begin{table*}[ht]
  \centering
  \caption{Fields recorded in a user-day subsession.}
  \label{tab:subsession-fields}
  \small
  \setlength{\tabcolsep}{4pt}
  \renewcommand{\arraystretch}{1.08}
  \begin{tabular*}{\textwidth}{@{\extracolsep{\fill}}p{0.27\textwidth}p{0.67\textwidth}@{}}
  \toprule
  Field & Meaning \\
  \midrule
  \texttt{subsession\_id} &
  Unique identifier composed of user ID and date, e.g.,
  \texttt{user\_role6::2026-03-16} \\
  \texttt{date} & Current episode date \\
  \texttt{user\_id} & Simulated user ID \\
  \texttt{profile\_snapshot} & User-profile snapshot before recommendation \\
  \texttt{candidate\_pool} & Candidate papers for the day \\
  \texttt{shown\_list} & Top-20 recommendation list shown by the system \\
  \texttt{system\_label} & Recommendation label generated by rules and scores \\
  \texttt{oracle\_label} & Offline evaluation label, used only for evaluation and log analysis \\
  \texttt{selected} &
  Whether the simulated user selected the paper; used only for behavior
  evaluation and later profile updates \\
  \texttt{drift\_state} & Current interest-drift state \\
  \texttt{reading\_signal\_state} & Recent reading-signal state \\
  \texttt{reading\_report} & Reading report for a selected paper \\
  \bottomrule
  \end{tabular*}
  \vspace{-2mm}
\end{table*}

The \texttt{oracle\_label} and \texttt{selected} fields are evaluation and logging fields. They are not used directly as model inputs during recommendation. At ranking time, the system can only use the user profile, paper metadata, paper content, explicit preferences, drift state, and historical behavioral signals. This avoids evaluation leakage and ensures that the metrics reflect recommendation ability.

\subsection{Recommendation Pipeline}

Each user-day episode is executed as follows:

\begin{enumerate}
\item Retrieve the candidate papers for the corresponding date from the fixed paper pool.
\item Compute the base relevance score from the user profile and paper content.
\item Check must-read author, institution, and keyword matches.
\item Apply interest-drift weighting to new interest topics and downweight suppressed old topics.
\item Apply recent reading-signal weights to short-term topics that repeatedly appear and are selected by the user.
\item Generate the Top-20 recommendation list.
\item Simulate user selections based on oracle labels, system rank, system label, and drift-topic matches.
\item Generate reading reports for selected papers and record token usage, episode metadata, and the drift timeline.
\end{enumerate}

This pipeline connects recommendation, selection, profile update, and reading assistance into a closed loop. The recommendation stage determines what the user sees. The selection stage simulates what the user actually clicks or reads. The profile-update stage adjusts short-term state based on selections. The reading-report stage produces deeper reading assistance for selected papers. PaperFlow therefore evaluates not only whether the ranking is correct, but also whether the system can maintain user interests, respond to interest changes, and provide explainable reading support during continuous use.

\subsection{Fixed Data and Fair Comparison}

All main and related comparison experiments use the same benchmark snapshot. The dataset is fixed for three reasons.

First, paper recommendation is highly sensitive to the candidate pool. If different models face different candidate papers, ranking differences may come from the data distribution rather than from the method. A fixed candidate pool ensures that all methods compete over the same papers.

Second, scientific papers are updated daily. If the experiment re-crawls data during evaluation, the runtime of different models would introduce external variation. A fixed historical paper pool avoids this source of bias.

Third, both interest drift and reading signals depend on temporal order. If the paper pool changes across runs, the candidate distribution on later dates may affect drift triggering and recovery. A fixed date range makes longitudinal behavior comparison reproducible.

Thus, all comparable results in the paper use the same fixed-pool setting. In the LLM comparison, only the language model used in recommendation and reading-report stages is replaced. The embedding model, user profiles, candidate pool, Top-20 budget, and metric computation remain fixed.
\subsection{Capability Coverage of Compared Methods}
\label{app:capability-coverage}

Table~\ref{tab:capability-coverage} summarizes which system capabilities are
covered by each compared method. The table is descriptive rather than an
evaluation metric; the main result tables evaluate whether these capabilities
improve recommendation quality under the same benchmark setting.

\begin{table*}[t]
  \centering
  \caption{Capability coverage of compared methods. A check mark denotes full
  coverage, a triangle denotes partial coverage, and a cross denotes absent
  coverage.}
  \label{tab:capability-coverage}
  \footnotesize
  \setlength{\tabcolsep}{2pt}
  \renewcommand{\arraystretch}{1.05}
  \providecommand{\stcapYes}{{\fontencoding{U}\fontfamily{pzd}\fontseries{m}\fontshape{n}\selectfont\char51}}
  \providecommand{\stcapPart}{$\triangle$}
  \providecommand{\stcapNo}{$\times$}
  \begin{tabular*}{\textwidth}{@{\extracolsep{\fill}}lcccccccccccc@{}}
  \toprule
  Method & CS & Prof. & Upd. & Drift & MR & Rank & Rpt. & Fb. & Wkly. & Impact & Behav. & Corr. \\
  \midrule
  Scholar Inbox
  & \stcapPart & \stcapYes & \stcapPart & \stcapNo & \stcapNo & \stcapYes
  & \stcapNo & \stcapNo & \stcapNo & \stcapNo & \stcapNo & \stcapNo \\
  Citation-Enhanced
  & \stcapNo & \stcapPart & \stcapNo & \stcapNo & \stcapNo & \stcapYes
  & \stcapNo & \stcapNo & \stcapNo & \stcapYes & \stcapNo & \stcapNo \\
  OMRC-MR
  & \stcapNo & \stcapNo & \stcapNo & \stcapNo & \stcapNo & \stcapYes
  & \stcapNo & \stcapNo & \stcapNo & \stcapNo & \stcapNo & \stcapNo \\
  UPR
  & \stcapPart & \stcapYes & \stcapNo & \stcapNo & \stcapNo & \stcapYes
  & \stcapNo & \stcapNo & \stcapNo & \stcapNo & \stcapNo & \stcapNo \\
  KUCNet
  & \stcapNo & \stcapPart & \stcapNo & \stcapNo & \stcapNo & \stcapYes
  & \stcapNo & \stcapNo & \stcapNo & \stcapPart & \stcapNo & \stcapNo \\
  PaperFlow
  & \stcapYes & \stcapYes & \stcapYes & \stcapYes & \stcapYes & \stcapYes
  & \stcapYes & \stcapYes & \stcapYes & \stcapYes & \stcapYes & \stcapYes \\
  \bottomrule
  \end{tabular*}
  \vspace{-2mm}
\end{table*}

The column abbreviations are: CS = cold start, Prof. = structured profile,
Upd. = profile update, Drift = interest-drift handling, MR = must-read rules,
Rank = ranked recommendation, Rpt. = reading report, Fb. = feedback use,
Wkly. = weekly or longitudinal use, Impact = citation or impact signal,
Behav. = behavioral signal, and Corr. = explicit correction support.
\section{User Profiles}
\label{app:user-profiles}

\subsection{Profile Design}

The 24 simulated users in PaperFlow cover diverse research directions, including GUI agents, AI for science, literature mining, embodied AI, multimodal reasoning, NLP, bioinformatics, protein structure, single-cell analysis, neuroscience, climate science, materials informatics, chemistry, high-energy physics, medical imaging, public health, agriculture, ocean science, psychology, economics, education research, astronomy, renewable energy, and science of science.

Each user profile contains four types of information: long-term interests, short-term interests, explicit preferences, and behavioral state. Long-term interests represent stable research directions. Short-term interests represent temporary topics formed by recent reading behavior. Explicit preferences represent must-read authors, institutions, or keywords. Behavioral state records interest drift and reading history.

A user profile is not a simple keyword set; it is a multi-layer structure. For example, an NLP user may have \texttt{nlp}, \texttt{large-language-model}, and \texttt{information-extraction} as core directions, and may also include secondary topics such as retrieval-augmented generation, long-context reasoning, and benchmark construction. The system must combine long-term directions, short-term behavior, and explicit rules to decide whether a paper is worth showing to the user.

\subsection{The 24 Simulated Researchers}

\begin{figure*}[t]
  \centering
  \makebox[\textwidth][c]{%
  \includegraphics[width=1.01\textwidth,trim=8 4 8 8,clip]{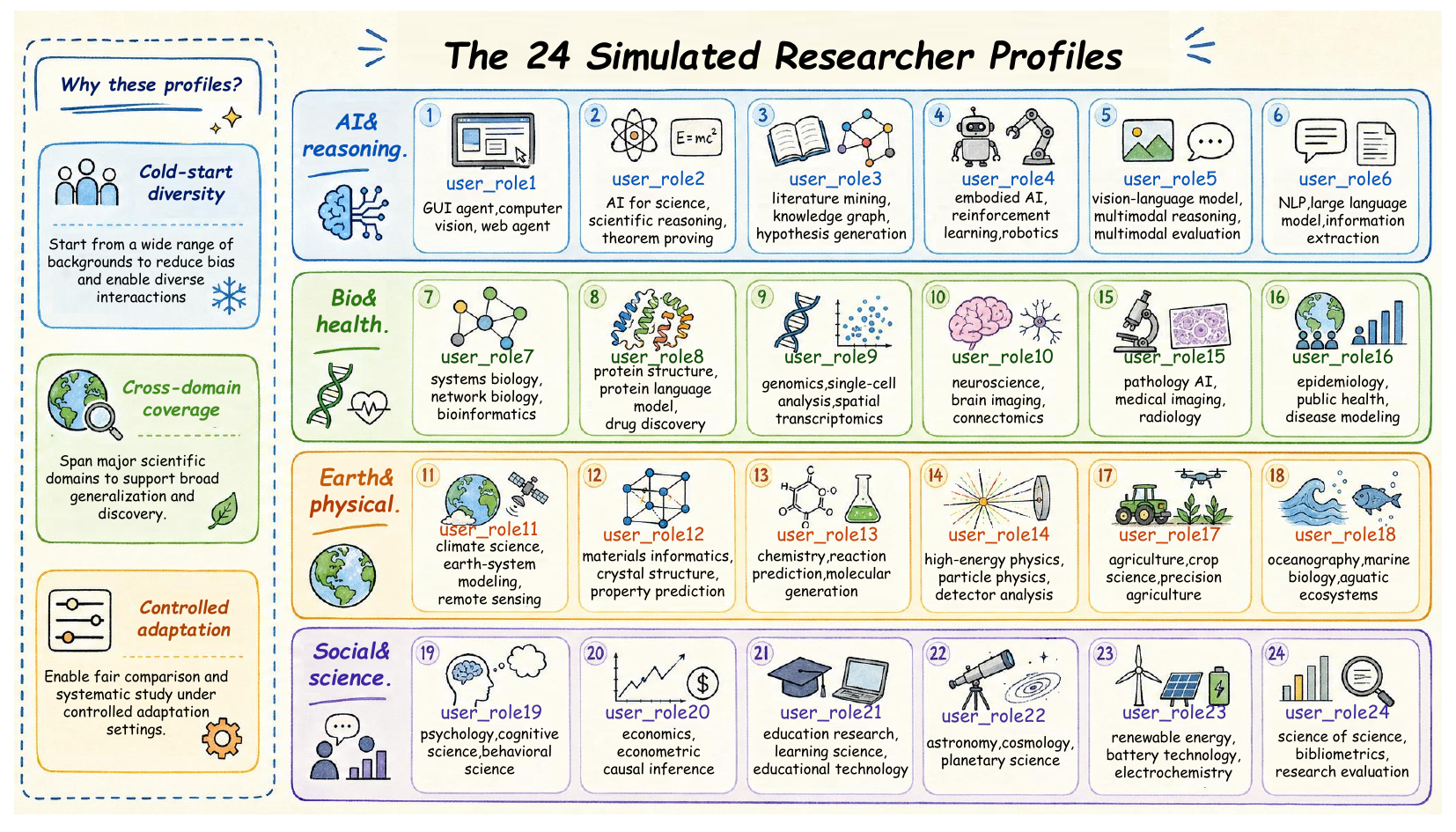}
  }
  \caption{The 24 simulated researcher profiles.}
  \label{fig:simulated-user-profiles}
  \vspace{-2mm}
\end{figure*}

Figure~\ref{fig:simulated-user-profiles} summarizes the 24 simulated users in PaperFlow. Each user has three main directions. The first direction has the largest weight, while the second and third directions create more specific cross-topic interests.

The goal of this user set is not to reproduce the full user distribution of a real platform. Instead, it creates a broad benchmark for testing cross-disciplinary recommendation ability. Compared with evaluation in a single field, the 24-user setup requires the system to handle AI, life science, physics, earth science, medicine, social science, and science-of-science topics, which is closer to the complexity of real scientific-paper recommendation.

\section{Hyperparameters}
\label{app:hyperparameters}

\subsection{Ranking Score Parameters}
\label{app:ranking-score-parameters}

The base PaperFlow ranking score combines semantic matching, profile matching,
preference signals, paper quality, drift adaptation, and recent reading
feedback. Let $I$, $T$, $H$, and $Q$ denote interest-vector similarity,
topic-weight match, author/institution heat, and paper-quality signal,
respectively. Let $B_m$, $B_d$, and $B_r$ denote the must-read, drift, and
reading-signal bonuses, and let $P_s$ denote the suppression penalty. The
ranking score is defined as
\begin{equation}
\begin{split}
S_{\mathrm{rank}} ={}&
0.35 I + 0.25 T + 0.20 H + 0.20 Q \\
&+ B_m + B_d + B_r - P_s .
\end{split}
\end{equation}

\begin{table*}[t]
  \centering
  \caption{Ranking-score parameters.}
  \label{tab:ranking-parameters}
  \small
  \setlength{\tabcolsep}{4pt}
  \renewcommand{\arraystretch}{1.08}
  \begin{tabular*}{\textwidth}{@{\extracolsep{\fill}}
    >{\raggedright\arraybackslash}p{0.34\textwidth}
    >{\centering\arraybackslash}p{0.08\textwidth}
    >{\raggedright\arraybackslash}p{0.52\textwidth}
    @{}}
  \toprule
  Parameter & Value & Description \\
  \midrule
  \texttt{w1\_interest\_vector} & 0.35 & Semantic similarity between paper content and the user's interest vector \\
  \texttt{w2\_topic\_weight} & 0.25 & Match between paper topics and user-profile topic weights \\
  \texttt{w3\_author\_\allowbreak institution} & 0.20 & Author, institution, and explicit-source preference weight \\
  \texttt{w4\_quality\_signal} & 0.20 & Paper-quality proxy weight \\
  \texttt{bonus\_must\_read} & 0.15 & Bonus for a must-read rule match \\
  \texttt{threshold\_high\_\allowbreak relevant} & 0.40 & Direct threshold for \texttt{high\_relevant} \\
  \texttt{threshold\_maybe\_\allowbreak interested} & 0.25 & Direct threshold for \texttt{maybe\_interested} \\
  \texttt{threshold\_edge\_\allowbreak relevant} & 0.15 & Threshold for \texttt{edge\_relevant} \\
  \texttt{min\_relevance\_signal} & 0.08 & Minimum personal relevance signal before a regular paper can enter the pushed candidate set \\
  \texttt{rank\_high\_fraction} & 0.10 & Top 10\% of relevant candidates can be promoted to \texttt{high\_relevant} if they reach the maybe threshold \\
  \texttt{rank\_maybe\_fraction} & 0.40 & Top 40\% of relevant candidates can be promoted to \texttt{maybe\_interested} if they reach the edge threshold \\
  \texttt{drift\_bonus\_shifting} & 0.08 & Weight for a new direction during drift shifting \\
  \texttt{drift\_bonus\_\allowbreak recovered} & 0.04 & Retained lightweight weight after drift recovery or stabilization \\
  \texttt{drift\_short\_topic\_\allowbreak bonus} & 0.03 & Additional weight for short-term drift-topic matches \\
  \texttt{reading\_signal\_\allowbreak short\_term\_\allowbreak bonus} & 0.05 & Additional weight for recent reading-signal matches \\
  \bottomrule
  \end{tabular*}
  \vspace{-2mm}
\end{table*}

We use four system labels:
\texttt{must\_read}, \texttt{high\_relevant},
\texttt{maybe\_interested}, and \texttt{edge\_relevant}.
The \texttt{must\_read} label is determined by explicit preference rules and has
higher priority than score thresholds, because it represents authors,
institutions, datasets, tasks, or keywords that the user explicitly wants to
track. Other labels are determined jointly by the total score, the personal
relevance floor, and rank-aware promotion. A regular paper must first satisfy
\texttt{min\_relevance\_signal}. It is then categorized by thresholds 0.40,
0.25, and 0.15. If a paper ranks near the top among relevant candidates, it can
be promoted to a higher display category even when its total score only reaches
a lower threshold.

\subsection{Explicit Preference Parameters}

Explicit preferences simulate rules that a user states directly, such as ``always show papers by this author,'' ``prioritize this institution,'' or ``track papers related to this benchmark or dataset.'' This module is intended to capture stable, high-confidence user preferences.

\begin{table*}[t]
  \centering
  \caption{Explicit preference types.}
  \label{tab:explicit-preferences}
  \small
  \setlength{\tabcolsep}{4pt}
  \renewcommand{\arraystretch}{1.08}
  \begin{tabular*}{\textwidth}{@{\extracolsep{\fill}}
    p{0.24\textwidth}
    p{0.35\textwidth}
    p{0.35\textwidth}
    @{}}
  \toprule
  Type & Example & Effect \\
  \midrule
  Author preference & A specified author or team & Raises ranking position after a match \\
  Institution preference & A specified university, lab, or company & Raises candidate priority after a match \\
  Keyword preference & long-context LLM, protein language model & Increases topic-match score after a match \\
  Task preference & information extraction benchmark, GUI grounding & Supports more fine-grained task matching \\
  Negative preference & not interested, downweight, remove, do not recommend & Downweights the corresponding topic \\
  \bottomrule
  \end{tabular*}
  \vspace{-2mm}
\end{table*}

Explicit preferences complement embedding similarity. Embeddings capture semantic relatedness, but may not express a user's strong rule about a specific author, institution, or task. Explicit preferences fill this gap. Conversely, relying only on explicit rules would miss semantically relevant papers without keyword matches. PaperFlow therefore combines both types of signals.

\subsection{Interest-Drift Parameters}
\label{app:interest_drift}
The interest-drift module simulates how a user's long-term interests change with reading behavior. The system first enters an \texttt{observing} state to monitor new-topic signals. When a new topic appears consecutively and passes the thresholds, the system locks an anchor topic and enters the \texttt{shifting} state. When the user returns to old core directions or the new and old directions become balanced, the system enters the \texttt{recovered} or \texttt{stable} state.

\begin{table*}[t]
  \centering
  \caption{Interest-drift parameters.}
  \label{tab:drift-parameters}
  \small
  \setlength{\tabcolsep}{4pt}
  \renewcommand{\arraystretch}{1.08}
  \begin{tabular*}{\textwidth}{@{\extracolsep{\fill}}
    >{\raggedright\arraybackslash}p{0.38\textwidth}
    >{\centering\arraybackslash}p{0.07\textwidth}
    >{\raggedright\arraybackslash}p{0.49\textwidth}
    @{}}
  \toprule
  Parameter & Value & Meaning \\
  \midrule
  \texttt{drift\_probability} & 0.5 & Probability of a drift opportunity in the simulation environment \\
  \texttt{ANCHOR\_SIGNAL\_WINDOW} & 3 & Sliding window for observing new-topic signals \\
  \texttt{ANCHOR\_REQUIRED\_\allowbreak CONSECUTIVE\_DAYS} & 2 & Number of consecutive days required for a new-topic signal \\
  \texttt{ANCHOR\_SIGNAL\_MIN\_HITS} & 2 & Minimum hits needed to lock a new topic \\
  \texttt{ANCHOR\_SIGNAL\_MIN\_RATIO} & 0.30 & Minimum share of the new topic among selected papers \\
  \texttt{ANCHOR\_SIGNAL\_MIN\_MARGIN} & 1 & Minimum hit margin over the second-highest topic \\
  \texttt{ANCHOR\_INTENT\_INCREMENT} & 0.15 & Intent-score increment when anchor evidence is observed \\
  \texttt{ANCHOR\_INTENT\_DECAY} & 0.05 & Intent-score decay when anchor evidence is insufficient \\
  \texttt{ANCHOR\_LOCK\_THRESHOLD} & 0.30 & Intent-score threshold for locking an anchor topic \\
  \texttt{ANCHOR\_PROGRESS\_STEP} & 0.40 & Drift-progress increment per update \\
  \texttt{ANCHOR\_SCORE\_STEP} & 0.24 & Drift-score increment per update \\
  \texttt{ANCHOR\_PRIMARY\_BOOST} & 0.12 & Primary-direction boost when the anchor is written into core directions \\
  \texttt{ANCHOR\_SECONDARY\_BOOST} & 0.06 & Secondary boost for topics related to the anchor \\
  \texttt{ANCHOR\_DOWNWEIGHT\_STEP} & 0.08 & Per-step downweighting amount for old directions during drift \\
  \texttt{ANCHOR\_MIN\_WEIGHT} & 0.05 & Minimum retained direction weight after downweighting \\
  \texttt{ANCHOR\_COMMITMENT\_DAYS} & 3 & Number of days to keep pushing the new direction after anchor lock \\
  \texttt{SIMULATION\_MAX\_\allowbreak DRIFT\_OPPORTUNITIES} & 5 & Maximum drift opportunities per user \\
  \texttt{SIMULATION\_\allowbreak CHECKFILE\_COOLDOWN\_\allowbreak EPISODES} & 8 & Cooldown episodes after completing one drift event \\
  \bottomrule
  \end{tabular*}
  \vspace{-2mm}
\end{table*}

A drift state does not directly imply higher recommendation quality. Its main purpose is to improve responsiveness to long-term interest changes. It may therefore introduce a small exploration cost on static oracle-based ranking metrics, while improving behavior consistency and long-term adaptation.

\subsection{Reading-Signal Parameters}

Reading signals represent short-term feedback formed by recent user selections. They are not the same as reading reports. The system updates short-term interest signals from the topics of recently selected papers and gives lightweight ranking bonuses to corresponding topics.

\begin{table*}[t]
  \centering
  \caption{Reading-signal parameters.}
  \label{tab:reading-signal-parameters}
  \small
  \setlength{\tabcolsep}{4pt}
  \renewcommand{\arraystretch}{1.08}
  \begin{tabular*}{\textwidth}{@{\extracolsep{\fill}}
    >{\raggedright\arraybackslash}p{0.42\textwidth}
    >{\centering\arraybackslash}p{0.07\textwidth}
    >{\raggedright\arraybackslash}p{0.45\textwidth}
    @{}}
  \toprule
  Parameter & Value & Meaning \\
  \midrule
  \texttt{PaperFlow\_READING\_SIGNAL\_\allowbreak WINDOW\_DAYS} & 21 & Recent reading-signal window \\
  \texttt{PaperFlow\_READING\_SIGNAL\_\allowbreak ACTIVATION\_COUNT} & 2 & Minimum number of selections required to activate a topic \\
  \texttt{PaperFlow\_READING\_SIGNAL\_\allowbreak TOPIC\_SEED\_WEAK} & 0.18 & Initial strength for a weak topic signal \\
  \texttt{PaperFlow\_READING\_SIGNAL\_\allowbreak TOPIC\_SEED\_STRONG} & 0.38 & Initial strength for a strong topic signal \\
  \texttt{PaperFlow\_READING\_SIGNAL\_\allowbreak TOPIC\_DELTA\_WEAK} & 0.03 & Weak topic increment \\
  \texttt{PaperFlow\_READING\_SIGNAL\_\allowbreak TOPIC\_DELTA\_STRONG} & 0.08 & Strong topic increment \\
  \texttt{PaperFlow\_READING\_SIGNAL\_\allowbreak CORE\_SEED\_STRONG} & 0.45 & Initial strength for a strong core-direction signal \\
  \texttt{PaperFlow\_READING\_SIGNAL\_\allowbreak CORE\_DELTA\_STRONG} & 0.08 & Strong core-direction increment \\
  \texttt{PaperFlow\_READING\_SIGNAL\_\allowbreak SHORT\_TERM\_BASE} & 0.35 & Base strength for short-term signals \\
  \texttt{PaperFlow\_READING\_SIGNAL\_\allowbreak SHORT\_TERM\_STEP} & 0.18 & Progress step for short-term signals \\
  \texttt{PaperFlow\_READING\_SIGNAL\_\allowbreak SHORT\_TERM\_STRONG\_\allowbreak BONUS} & 0.22 & Additional bonus for strong short-term signals \\
  \bottomrule
  \end{tabular*}
  \vspace{-2mm}
\end{table*}

Reading signals are intentionally lightweight. They should not overwrite the long-term profile or amplify one accidental selection into a long-term interest change. The system therefore uses a window, an activation count, and weak/strong signal distinctions. A topic becomes a short-term preference only after it has been selected multiple times in the recent window.

\subsection{Reading-Report Parameters}

The reading-report module performs PDF- or metadata-driven structured analysis for selected papers. Report generation prefers parsed PDF content and semantic-retrieval evidence. If PDF retrieval fails, the system falls back to a simplified analysis based on title, abstract, and metadata.

\begin{table}[t]
  \small
  \centering
  \vspace{-1mm}
  \caption{Reading-report parameters.}
  \vspace{-1mm}
  \label{tab:reading-report-parameters}
  \setlength{\tabcolsep}{1.5mm}
  \begin{tabular*}{\columnwidth}{@{\extracolsep{\fill}}p{0.62\columnwidth}p{0.25\columnwidth}@{}}
  \toprule
  Parameter & Value \\
  \midrule
  \texttt{READING\_REPORT\_PDF\_TIMEOUT} & 60 seconds \\
  \texttt{READING\_REPORT\_ARXIV\_TIMEOUT} & 12 seconds \\
  \texttt{READING\_REPORT\_ABSTRACT\_CHARS} & 1200 \\
  \texttt{READING\_REPORT\_SECTION\_CHARS} & 1800 \\
  \texttt{READING\_REPORT\_CHUNK\_CHARS} & 1200 \\
  \texttt{READING\_REPORT\_CHUNK\_OVERLAP} & 180 \\
  \texttt{READING\_REPORT\_EVIDENCE\_TOP\_K} & 3 \\
  \texttt{READING\_REPORT\_PROFILE\_\allowbreak RETRIEVAL\_WEIGHT} & 0.25 \\
  \texttt{READING\_REPORT\_LLM\_MAX\_TOKENS} & 4096 \\
  \bottomrule
  \end{tabular*}
  \vspace{-3mm}
\end{table}

A reading report is not part of the ranking metric. It is a post-recommendation reading-assistance capability. It converts selected papers into structured reading material, including a one-sentence summary, research background, core method, key results, contributions, limitations, relevance to the user's research, and suggested reading focus. For researchers, the recommendation list identifies what to read first; the reading report helps decide how and why to read it.

\subsection{LLM and Embedding API Setting}

In the model-comparison experiment, the LLM API and embedding API are configured separately. This ensures that the LLM comparison changes only the language model used in generation, parsing, or judging stages, while leaving candidate-paper embeddings unchanged. Both the main experiment and model comparison use \texttt{Qwen3-Embedding-8B} as the embedding model.

\begin{table}[t]
  \small
  \centering
  \vspace{-1mm}
  \caption{API and output settings for model comparison.}
  \vspace{-1mm}
  \label{tab:api-settings}
  \setlength{\tabcolsep}{1.5mm}
  \begin{tabular*}{\columnwidth}{@{\extracolsep{\fill}}p{0.32\columnwidth}p{0.58\columnwidth}@{}}
  \toprule
  Module & Setting \\
  \midrule
  LLM API & Configured separately for each experimental model, e.g., Gemini, GPT, Qwen, DeepSeek, and Xiaomi MiMo \\
  Embedding API & Separately configured and fixed to \texttt{Qwen3-Embedding-8B} \\
  Token records & Daily records of embedding tokens, LLM tokens, total tokens, and call count \\
  Output directory & Each model writes to an independent results directory \\
  Database & Before each full run, benchmark state is reset while the papers table is retained \\
  \bottomrule
  \end{tabular*}
  \vspace{-3mm}
\end{table}

This design avoids confounds in model comparison. If different LLMs also used different embedding models, performance differences could come from embeddings rather than LLM reasoning or parsing. Fixing the embedding model makes the comparison closer to replacing the LLM within the same recommendation framework.

\section{Evaluation Metrics}
\label{app:evaluation-metrics}

\subsection{Main Metric Suite}

The main experiment and related comparisons use the same recommendation metrics so that methods can be compared directly. The metrics cover ranking quality, recall, behavioral consistency, and aggregate recommendation quality.

\begin{figure*}[t]
  \centering
  \includegraphics[width=\textwidth,trim=20 10 60 10,clip]{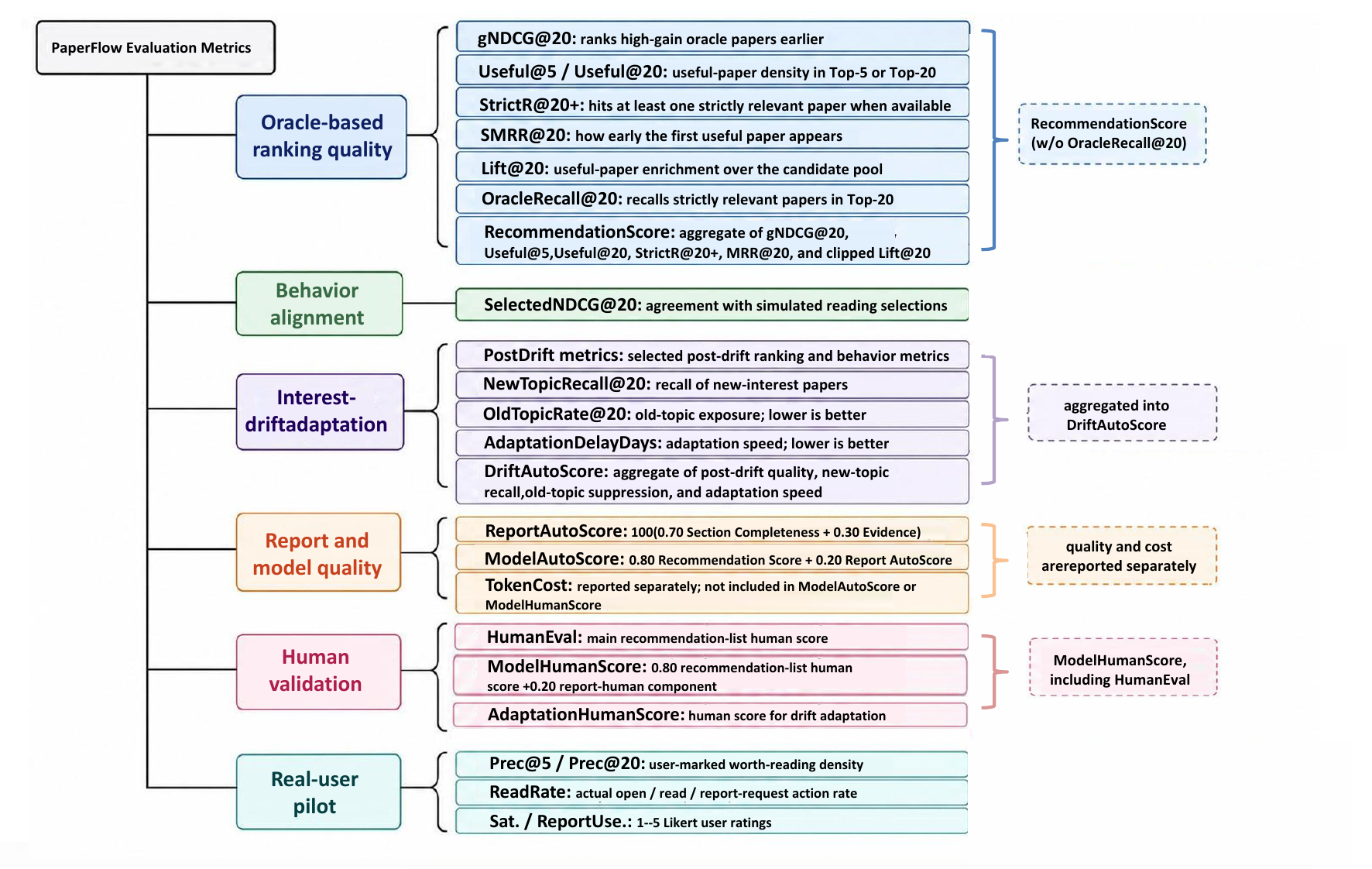}
  \caption{Overview of PaperFlow evaluation metrics. The figure groups metrics
  by oracle-based ranking quality, behavior alignment, interest-drift adaptation,
  report and model quality, human validation, and real-user pilot evaluation.
  Composite scores aggregate the indicated automatic metrics, while separately
  reported diagnostics, human scores, token cost, and real-user pilot metrics are
  not used as inputs to the main automatic recommendation score.}
  \label{fig:metric-overview}
  \vspace{-2mm}
\end{figure*}

Figure~\ref{fig:metric-overview} summarizes the metric families and their
aggregation relationships before we define each metric individually.

\begin{table}[t]
  \small
  \centering
  \vspace{-1mm}
  \caption{Main recommendation metrics.}
  \vspace{-1mm}
  \label{tab:main-metrics}
  \setlength{\tabcolsep}{1.5mm}
  \begin{tabular*}{\columnwidth}{@{\extracolsep{\fill}}p{0.38\columnwidth}p{0.52\columnwidth}@{}}
  \toprule
  Metric & Meaning \\
  \midrule
  \texttt{gNDCG@20} & Top-20 ranking quality computed from oracle gains \\
  \texttt{Useful@5} & Fraction of useful papers in the Top-5 \\
  \texttt{Useful@20} & Fraction of useful papers in the Top-20 \\
  \texttt{OracleRecall@20} & Recall of strictly relevant papers in the Top-20 \\
  \texttt{Lift@20} & Top-20 useful rate divided by candidate-pool useful rate \\
  \texttt{StrictR@20+} & Whether a strictly relevant paper is recalled in the Top-20 when such papers exist \\
  \texttt{MRR@20} & Reciprocal rank of the first positive item in the Top-20 \\
  \texttt{SelectedNDCG@20} & Top-20 NDCG using simulated user selection as gain \\
  \texttt{RecommendationScore} & Aggregate recommendation score \\
  \bottomrule
  \end{tabular*}
  \vspace{-3mm}
\end{table}

\subsection{\texttt{gNDCG@20}}

\texttt{gNDCG@20} measures the overall ranking quality of the Top-20 list. Similar to standard NDCG, it checks whether highly relevant papers are ranked earlier. We convert oracle labels into gains: \texttt{strong\_relevant}, \texttt{relevant}, \texttt{weak\_relevant}, and \texttt{irrelevant} correspond to 2.0, 1.0, 0.5, and 0.0, respectively.

This metric evaluates whether the system concentrates valuable papers near the front of a limited display budget. In paper recommendation, users usually do not browse the entire candidate pool, but focus on the first few items. Thus, even if the Top-20 contains useful papers, usefulness decreases if those papers are placed too late.

\subsection{\texttt{Useful@5} and \texttt{Useful@20}}

\texttt{Useful@5} is the fraction of papers in the Top-5 whose oracle label is \texttt{strong\_relevant}, \texttt{relevant}, or \texttt{weak\_relevant}. It focuses on recommendation density at the very top of the list.

\texttt{Useful@20} is the fraction of useful papers in the Top-20. It measures the information density of the full displayed list. Compared with \texttt{Useful@5}, \texttt{Useful@20} better captures how many potentially useful papers are covered within the display budget.

Both metrics are needed because researchers may either skim only the first few papers or inspect the full Top-20. \texttt{Useful@5} is stricter and emphasizes head-of-list quality, while \texttt{Useful@20} is broader and emphasizes overall display quality.

\subsection{\texttt{OracleRecall@20}}

\texttt{OracleRecall@20} measures the fraction of strictly relevant papers recalled in the Top-20. Strict relevance typically includes \texttt{strong\_relevant} and \texttt{relevant}, but not \texttt{weak\_relevant}. This metric evaluates whether the system misses papers that are truly important to the user.

In scientific recommendation, missing one highly relevant paper can be more harmful than recommending several weakly relevant ones. This is especially true for must-read authors, key benchmarks, or highly matched research directions. \texttt{OracleRecall@20} is therefore an important coverage metric.

\subsection{\texttt{Lift@20}}

\texttt{Lift@20} is the Top-20 useful rate divided by the useful rate of the entire candidate pool. If useful papers are rare in the candidate pool but the system concentrates more of them in the Top-20, \texttt{Lift@20} becomes high.

This metric measures the gain of recommendation over randomly browsing the candidate pool. Since the average daily candidate pool contains more than 400 papers and the user sees only 20, the system's core value is to extract a higher-density set of useful papers. \texttt{Lift@20} directly reflects this filtering ability.

\subsection{\texttt{StrictR@20+}}

\texttt{StrictR@20+} measures whether the system recalls at least one strictly relevant paper in the Top-20 for episodes where strictly relevant papers exist. It is a recall- and usability-oriented metric that asks whether the system gives the user at least one truly important paper in key scenarios.

This metric is especially important for long-tail episodes. On some days, only a few strictly relevant papers may appear in the candidate pool. If the system misses all of them, the user experience for that day drops sharply. \texttt{StrictR@20+} captures this failure mode.

\subsection{\texttt{MRR@20}}

\texttt{MRR@20} computes the reciprocal rank of the first positive item within the Top-20. If the first recommendation is useful, the episode MRR is 1. If the first useful paper appears at rank 5, the MRR is 1/5.

MRR asks how quickly the user encounters the first useful recommendation. In a real system, earlier useful items help users trust the system and continue using it. This metric complements NDCG and useful-rate metrics.

\subsection{\texttt{SelectedNDCG@20}}

\texttt{SelectedNDCG@20} uses simulated user selection behavior as gain and measures agreement between ranking and the papers that the user actually selects. If selected papers are ranked early, the score is high; if selected papers appear late, the score is low.

This metric differs from oracle-label metrics. Oracle labels measure static relevance to the user profile, while selection behavior reflects what the user might actually read in the current context. They need not perfectly agree. For example, a paper may be labeled weakly relevant by the oracle, but may still be selected because of recent reading signals or interest drift. \texttt{SelectedNDCG@20} therefore captures dynamic behavioral fit.

\subsection{\texttt{RecommendationScore}}
\label{sec:appendix-rec-score}

RecommendationScore is the aggregate quality score of recommendations reported in
the main result tables. Let $G$, $U_5$, $U_{20}$, $Q$, $M$, and $L$ denote
gNDCG@20, Useful@5, Useful@20, StrictR@20+, MRR@20, and Lift@20, respectively.
We define RecommendationScore $R_s$ as
\begin{equation}
\label{eq:recommendation-score}
\begin{aligned}
R_s = 100(&0.25G + 0.15U_5 + 0.15U_{20} \\
          &+ 0.20Q + 0.15M \\
          &+ 0.10\min(L/15, 1)).
\end{aligned}
\end{equation}
The Lift@20 term is clipped by $\min(L/15, 1)$ to prevent extremely large
lift values from dominating the aggregate score.

The score is used only as a compact display metric. It does not replace the
individual metrics. We therefore interpret RecommendationScore together with
the component ranking metrics, SelectedNDCG@20, and human evaluation. In
particular, a high RecommendationScore does not imply that a method is better
in every dimension; it mainly summarizes oracle-based ranking quality under a
fixed Top-20 recommendation budget.

\subsection{\texttt{ReportAutoScore}}
\label{sec:appendix-report-auto-score}

\texttt{ReportAutoScore} is an automatic proxy for reading-report quality. It
measures whether the generated report is structurally complete and supported
by evidence anchors. Let $C_s$ and $E_c$ denote SectionCompleteness and
EvidenceCoverage, respectively. We define ReportAutoScore $R_a$ as
\begin{equation}
\label{eq:report-auto-score}
R_a = 100(0.70C_s + 0.30E_c).
\end{equation}
\texttt{SectionCompleteness} measures whether the required report fields are
filled, and \texttt{EvidenceCoverage} measures whether the report contains
retrieved evidence, evidence anchors, or full-text parsing support.

\subsection{\texttt{ModelAutoScore}}
\label{sec:appendix-model-auto-score}

The automatic model-quality score $M_a$ combines recommendation quality and
report quality, while token cost is reported separately:
\begin{equation}
\label{eq:model-auto-score}
M_a = 0.80R_s + 0.20R_a.
\end{equation}
Because \texttt{ReportAutoScore} already includes both structure completeness
and evidence coverage, these sub-metrics are not added again to
\texttt{ModelAutoScore}.

\subsection{\texttt{DriftAutoScore}}
\label{sec:appendix-drift-auto-score}

\texttt{DriftAutoScore} ($D_s$) summarizes adaptation-oriented signals into a
0--100 aggregate. Let $O$ denote OldTopicRate@20 and $D$ denote
AdaptationDelayDays. We first define old-topic suppression $S_o$ and
normalized delay score $S_d$:
\begin{equation}
\label{eq:drift-aux-scores}
S_o = 1 - O, \qquad
S_d = 1 - \frac{D - D_{\min}}{D_{\max} - D_{\min}}.
\end{equation}
Let $G_d$, $U_d$, $R_d$, $B_d$, and $N_d$ denote PostDrift gNDCG@20,
PostDrift Useful@20, PostDrift OracleRecall@20, PostDrift SelectedNDCG@20,
and NewTopicRecall@20, respectively. We define DriftAutoScore $D_s$ as
\begin{equation}
\label{eq:drift-auto-score}
\begin{aligned}
D_s = 100(&0.10G_d + 0.10U_d + 0.10R_d \\
          &+ 0.20B_d + 0.25N_d \\
          &+ 0.15S_o + 0.10S_d).
\end{aligned}
\end{equation}
The score weights post-drift ranking quality, behavioral alignment with
post-drift selections, recall of new-topic papers, suppression of residual
old-topic exposure, and adaptation speed. It is intended to compare
adaptation behavior on drift episodes rather than to replace static
oracle-based ranking metrics.

\subsection{Relationship Among Metrics}

We use both oracle-based and behavior-based metrics because scientific recommendation has two different goals. The first is to rank statically relevant papers near the top according to predefined relevance criteria. The second is to predict what the user will actually choose at the current stage of continuous reading.

Oracle-based metrics include \texttt{gNDCG@20}, \texttt{Useful@5}, \texttt{Useful@20}, \texttt{OracleRecall@20}, \texttt{Lift@20}, \texttt{StrictR@20+}, and \texttt{MRR@20}. They are suited for comparing ranking ability under fixed labels. The main behavior-based metric is \texttt{SelectedNDCG@20}, which evaluates agreement with simulated user behavior.

The two metric families can conflict slightly. For example, removing the drift module may make the system more conservative and prioritize papers aligned with stable long-term profiles, improving oracle-based ranking metrics. Full PaperFlow, however, retains drift exploration and may better match subsequent user selections, producing higher \texttt{SelectedNDCG@20}. This is not a metric bug; it reflects the trade-off between static relevance and dynamic behavioral adaptation.

\section{Prompt Templates}
\label{app:prompt-templates}

\subsection{How Prompts Are Used}

Prompts in PaperFlow are mainly used for structured parsing and reading-report generation, rather than directly asking an LLM to produce recommendations from evaluation labels. Recommendation ranking is determined by user profiles, paper content, rule-based signals, and structured fields produced by the model. Publishing the prompt templates improves reproducibility and reduces ambiguity caused by unstable LLM output formats.

We do not design a separate prompt for a subsession. A subsession is a logging and contextual unit, while prompts are called for profile parsing, direction extraction, recommendation-explanation generation, and reading-report generation.

\subsection{User-Profile Parsing Prompt}

This prompt parses a user's natural-language description of research interests and converts it into structured profile-update signals.

\begin{promptbox}[User-Profile Parsing Prompt]
\begin{lstlisting}
System:
You are an academic profile parsing assistant. Your task is to parse
a user's description of research interests and extract structured
information.

Pay special attention to negation, correction, contrast, and modal
phrases. If the user expresses "not interested", "downweight",
"remove", "do not recommend", or similar meanings, treat it as a
negative adjustment rather than a positive interest.

Phrases such as "GUI Agent", "Cold Start", and "protein language
model" must be kept as complete topics. Do not split them into
separate words.

User:
{user_input}

Return JSON only:
{
  "action": "adjust_interest | adjust_weight | add_must_read |
             remove_must_read | unknown",
  "direction": "increase | decrease | null",
  "topics": ["recognized research directions or topics"],
  "confidence": 0.0,
  "reasoning": "brief explanation"
}
\end{lstlisting}
\end{promptbox}

The key requirement is to recognize negative expressions and preserve complete phrases. Many scientific interests are fixed multi-word concepts, such as \texttt{protein language model}, \texttt{world model}, and \texttt{GUI agent}. Incorrectly splitting them would harm subsequent topic matching.

\subsection{Research-Direction Extraction Prompt}

This prompt extracts up to three research directions from a user's self-description and prioritizes specific topics.

\begin{promptbox}[Research-Direction Extraction Prompt]
\begin{lstlisting}
System:
Extract up to 3 research directions from a user's self-description.
Keep multi-word phrases intact and prefer specific topics over broad
umbrellas.
Do not split phrases like "protein language model" or
"world model for epidemiology".
Known directions include: {known_direction_list}.
Return JSON only.

User:
{user_description}

Return:
{
  "directions": [
    {
      "name": "english-kebab-case-or-normalized-name",
      "display_name": "human-readable display name",
      "confidence": 0.0,
      "source_text": "matched phrase from the user text",
      "is_known": true
    }
  ]
}
\end{lstlisting}
\end{promptbox}

The purpose is to normalize natural-language interest descriptions into computable directions. The normalized directions are used for profile construction, topic-weight computation, and embedding-query construction.

\subsection{Recommendation-Explanation Prompt}

This prompt generates a short explanation for papers in the Top-20 list. The explanation is not the final ranking criterion; it explains why the system considers a paper relevant.

\begin{promptbox}[Recommendation-Explanation Prompt]
\begin{lstlisting}
System:
You are an assistant for explaining scientific-paper recommendations.
Based on the user profile, paper metadata, system label, and matching
signals, generate a concise, specific explanation that is faithful to
the input information.

Requirements:
1. Do not invent methods, experiments, or conclusions that do not
   appear in the paper information.
2. If the paper is only topically related, explicitly say it is
   topically related. Do not exaggerate it as directly solving the
   user's problem.
3. If a must-read author, institution, or keyword is matched, state
   the reason.
4. If the paper is related to an interest-drift direction, state the
   corresponding new interest topic.
5. Output only one or two sentences.

User:
{
  "user_profile": {
    "core_directions": {...},
    "must_read": {...},
    "drift_state": "...",
    "reading_signal_topics": [...]
  },
  "paper": {
    "title": "...",
    "abstract": "...",
    "authors": [...],
    "institutions": [...],
    "keywords": [...]
  },
  "signals": {
    "system_label": "...",
    "score": 0.0,
    "matched_topics": [...],
    "matched_must_read": [...],
    "drift_matches": [...]
  }
}

Return JSON only:
{
  "reason": "recommendation explanation",
  "faithfulness_note": "which input fields support this explanation"
}
\end{lstlisting}
\end{promptbox}

The prompt emphasizes faithfulness and restraint. Over-generated explanations can create false expectations. This is especially harmful in scientific-paper recommendation, where a system must avoid exaggerating ``topic similarity'' into ``direct methodological relevance'' or ``immediately usable experimental findings.''

\subsection{Reading-Report Generation Prompt}

This prompt generates a structured reading report for a selected paper. The input includes paper metadata, abstract, PDF section snippets, user profile, a heuristic draft, and semantic-retrieval evidence.

\begin{promptbox}[Reading-Report Generation Prompt]
\begin{lstlisting}
System:
You are a scientific-paper reading assistant. Based on the provided
paper metadata, abstract, available section excerpts, and user profile,
generate a structured reading report suitable for a research document.

If retrieved_evidence is provided, prioritize the evidence snippets
retrieved from the PDF. Then use heuristic_draft for polishing and
completion.

When retrieved_evidence conflicts with heuristic_draft, prioritize
retrieved_evidence.

Do not invent specific experimental numbers. If information is
insufficient, be conservative and state that the original paper should
be checked.

Output only a JSON object. Do not output Markdown, explanations, or
code blocks.

User:
{
  "paper": {
    "title": "...",
    "authors": ["..."],
    "abstract": "...",
    "venue": "...",
    "publish_date": "...",
    "arxiv_id": "...",
    "doi": "..."
  },
  "sections": {
    "introduction": "...",
    "method": "...",
    "results": "...",
    "discussion": "...",
    "conclusion": "..."
  },
  "user_profile": {
    "top_directions": ["..."],
    "methodology_preferences": {...},
    "report_preferences": {...}
  },
  "heuristic_draft": {...},
  "retrieved_evidence": {...},
  "field_evidence_map": {...}
}

Return:
{
  "one_sentence_summary": "...",
  "research_background": "...",
  "core_method": "...",
  "key_results": "...",
  "main_contributions": ["..."],
  "limitations": ["..."],
  "relevance_points": ["..."],
  "reading_focus": ["..."],
  "recommendation_label": "Strongly recommended | Recommended |
                           Worth skimming | Read if needed",
  "analysis_note": "..."
}
\end{lstlisting}
\end{promptbox}

The key design goals are evidence priority and hallucination prevention. The system first uses evidence retrieved from the PDF. If PDF retrieval fails, it falls back to a simplified report based on title, abstract, and metadata. Experimental numbers, dataset results, and conclusions must come from the input evidence rather than from model completion.

\subsection{Reading-Report Markdown Template}

\begin{promptbox}[Reading-Report Markdown Template]
\begin{lstlisting}
{paper_title}

Basic Information

Authors: {authors}
Institutions: {institutions}
Source: {source}
Date: {publish_date}
Recommendation level: {recommendation_label}
Estimated reading time: {estimated_reading_minutes} minutes
Analysis source: {analysis_source}

One-Sentence Summary

{one_sentence_summary}

Research Background

{research_background}

Core Method

{core_method}

Key Results

{key_results}

Contributions

{contribution_1}
{contribution_2}
{contribution_3}

Limitations

{limitation_1}
{limitation_2}

Relation to My Research

{relevance_point_1}
{relevance_point_2}

How to Read It

{reading_focus_1}
{reading_focus_2}

PDF Evidence Anchors

Background evidence: {background_evidence}
Method evidence: {method_evidence}
Result evidence: {result_evidence}

Code and Resources

Code: {code_url_or_na}
Data: {data_url_or_na}
Project page: {project_url_or_na}
Paper: {paper_url}

My Notes

Point I most want to reproduce or borrow:
Connection to my current research:
Whether to follow the author, code, or follow-up work:
\end{lstlisting}
\end{promptbox}

This template can be pasted directly into a research document. It separates paper metadata, content understanding, user relevance, and next actions, allowing the user not only to understand the paper but also to decide whether to read it deeply, reproduce it, or track follow-up work.

\subsection{Human-Evaluation Prompts}

The human-evaluation prompts correspond to the three human-score definitions in Appendix~\ref{sec:human-eval-protocol}. The main experiment evaluates only recommendation quality. The model comparison evaluates both recommendation and reading-report quality. The drift-specific evaluation evaluates only adaptation after interest change. All prompts use a 1--5 Likert scale. Under blind evaluation, method name, oracle label, selected field, and system score are hidden.

\paragraph{Main listwise recommendation-quality prompt.}

\begin{promptbox}[Main Listwise Recommendation-Quality Prompt]
\begin{lstlisting}
You will see a researcher profile and an anonymized Top-20 recommendation
list for one user-day episode. Please evaluate the whole list, not only a
single paper.

Use a 1-5 Likert scale:
1 = very poor
2 = poor
3 = fair
4 = good
5 = very good

Evaluate the following dimensions:
1. ProfileMatch: whether the list matches the user's profile and current
   research interests.
2. RankingQuality: whether stronger or more useful papers appear earlier
   in the Top-20 list.
3. DecisionUsefulness: whether the list helps the user decide what to read,
   skim, or skip.
4. DiversityFocusBalance: whether the list balances focused relevance with
   useful breadth.

Return JSON:
{
  "ProfileMatch": 1,
  "RankingQuality": 1,
  "DecisionUsefulness": 1,
  "DiversityFocusBalance": 1,
  "comments": "brief rationale"
}
\end{lstlisting}
\end{promptbox}

\paragraph{Model-comparison report prompt.}

\begin{promptbox}[Model-Comparison Report Prompt]
\begin{lstlisting}
You will see a researcher profile, a recommended paper, the paper
abstract, a system recommendation explanation, and a reading report.
Please evaluate recommendation quality and report quality separately.

Use a 1-5 Likert scale and return JSON:
{
  "HumanRelevance": 1,
  "HumanUsefulness": 1,
  "RecommendationDecisionHelpfulness": 1,
  "ReportFaithfulness": 1,
  "ReportSpecificity": 1,
  "ReportDecisionHelpfulness": 1,
  "comments": "brief rationale"
}
\end{lstlisting}
\end{promptbox}

\paragraph{Interest-drift prompt.}

\begin{promptbox}[Interest-Drift Prompt]
\begin{lstlisting}
You will see a user profile with interest drift, the interest directions
before and after the drift, and recommended papers after the drift.
Please evaluate whether the recommendation reasonably adapts to the
user's new interests.

Use a 1-5 Likert scale and return JSON:
{
  "NewTopicFit": 1,
  "AdaptationAppropriateness": 1,
  "OldNewBalance": 1,
  "DriftDecisionHelpfulness": 1,
  "comments": "brief rationale"
}
\end{lstlisting}
\end{promptbox}

\FloatBarrier
\section{Human Evaluation Protocol}
\label{sec:human-eval-protocol}
\label{app:human-evaluation-protocol}

This section defines the human evaluation protocol. To avoid mixing human scores from different tasks, we define separate scores for main-experiment recommendation quality, LLM comparison, and interest-drift adaptation. The main paper reports only the corresponding human-score columns; all automatic-human correlation plots are placed in the appendix.

In this paper, \emph{blind human evaluation} means that annotators are shown only the information needed for the judgment task, such as the user profile, anonymized recommendation list, and, for model comparison, sampled reading-report excerpts. They are not shown method or model identities, oracle relevance labels, simulated selections, system-side scores, or automatic metric values. The protocol is metric-aligned in the sense that the human dimensions evaluate the same recommendation, report, or drift-adaptation constructs measured by the automatic scores, but annotators do not see those scores and their judgments are not used to define the simulator labels.
\subsection{Main-Experiment Human Evaluation: \texttt{HumanEval}}

In the main experiment, \texttt{HumanEval} denotes a blind listwise human
evaluation score over anonymized Top-20 recommendation lists. It measures
ordinary recommendation quality only and is not used as human-labeled relevance
ground truth. Each annotation unit is one method--episode Top-20 list.

Let $H_p$, $H_r$, and $H_d$ denote ProfileMatch, RankingQuality, and
DecisionUsefulness, respectively. We define the main HumanEval score as
\begin{equation}
\label{eq:human-eval}
H_e = 20 \cdot \operatorname{mean}(H_p, H_r, H_d).
\end{equation}
We additionally collect DiversityFocusBalance as a diagnostic dimension, but
do not include it in the primary HumanEval score.

\begin{table}[t]
\centering
\caption{Human dimensions for main-experiment listwise recommendation quality.}
\label{tab:human-eval-dimensions}
\small
\resizebox{\linewidth}{!}{%
\begin{tabular}{p{0.42\linewidth}p{0.48\linewidth}}
\hline
Dimension & Meaning \\
\hline
\texttt{ProfileMatch} & Whether the list matches the user's profile and current research interests \\
\texttt{RankingQuality} & Whether stronger or more useful papers appear earlier in the Top-20 list \\
\texttt{DecisionUsefulness} & Whether the list helps the user decide what to read, skim, or skip \\
\texttt{DiversityFocusBalance} & Whether the list balances focused relevance with useful breadth \\
\hline
\end{tabular}%
}
\end{table}

For the main human evaluation, we sample six user-day episodes and score six
methods per episode, yielding 36 anonymized method--episode Top-20 lists. Three
annotators independently score each list. Annotators see the user profile and
the anonymized recommendation list, but not the method name, oracle labels,
simulated selections, system scores, or automatic metric values.

\subsection{Model-Comparison Human Evaluation: \texttt{ModelHumanScore}}

The LLM comparison evaluates both recommendation quality and reading-report
quality, so we define \texttt{ModelHumanScore}. Let $H_r^{\mathrm{rec}}$,
$H_u^{\mathrm{rec}}$, and $H_d^{\mathrm{rec}}$ denote HumanRelevance,
HumanUsefulness, and RecommendationDecisionHelpfulness, respectively. The
recommendation component $M_h^{\mathrm{rec}}$ is defined as
\begin{equation}
\label{eq:human-recommendation-score}
M_h^{\mathrm{rec}}
= 20 \cdot \operatorname{mean}
(H_r^{\mathrm{rec}}, H_u^{\mathrm{rec}}, H_d^{\mathrm{rec}}).
\end{equation}

Let $H_f^{\mathrm{rep}}$, $H_s^{\mathrm{rep}}$, and $H_d^{\mathrm{rep}}$
denote ReportFaithfulness, ReportSpecificity, and ReportDecisionHelpfulness,
respectively. We define the report component $M_h^{\mathrm{rep}}$ as
\begin{equation}
\label{eq:report-human-score}
M_h^{\mathrm{rep}}
= 20 \cdot \operatorname{mean}
(H_f^{\mathrm{rep}}, H_s^{\mathrm{rep}}, H_d^{\mathrm{rep}}).
\end{equation}

The final \texttt{ModelHumanScore} $M_h$ combines recommendation and report
quality:
\begin{equation}
\label{eq:model-human-score}
M_h = 0.80M_h^{\mathrm{rec}} + 0.20M_h^{\mathrm{rep}}.
\end{equation}
It is also normalized to a 0--100 scale.

The recommendation-list component uses three dimensions: relevance to the user
profile, usefulness of the ranked list, and helpfulness for deciding what to
read. The reading-report component uses three dimensions: faithfulness to the
paper information, specificity rather than genericness, and helpfulness for the
reading decision. We sample six common user-day episodes across the 14 completed
LLM backbones, yielding 84 anonymized model--episode list-level samples. Three
annotators score each sample, producing 18 annotations per model. Reviewers see
the user profile, the anonymized Top-10 recommendation list, and up to two
reading-report excerpts, but not the model name, oracle labels, selected fields,
system scores, or automatic metric values.

\subsection{Drift-Specific Human Evaluation: \texttt{AdaptationHumanScore}}

To test whether the interest-drift mechanism genuinely adapts to changing user interests, we define \texttt{AdaptationHumanScore}. This score is used only for episodes involving interest drift and is not mixed with ordinary \texttt{HumanEval}. Let $A_n$, $A_p$, $A_b$, and $A_d$ denote NewTopicFit, AdaptationAppropriateness, OldNewBalance, and DriftDecisionHelpfulness, respectively. We define AdaptationHumanScore $A_h$ as
\begin{equation}
\label{eq:adaptation-human-score}
A_h = 20 \cdot \operatorname{mean}(A_n, A_p, A_b, A_d).
\end{equation}
It is normalized to a 0--100 scale and used only for drift episodes.

\begin{table}[t]
  \small
  \centering
  \vspace{-1mm}
  \caption{Human dimensions for drift-specific adaptation.}
  \vspace{-1mm}
  \label{tab:adaptation-human-dimensions}
  \setlength{\tabcolsep}{1.5mm}
  \begin{tabular*}{\columnwidth}{@{\extracolsep{\fill}}p{0.50\columnwidth}p{0.40\columnwidth}@{}}
  \toprule
  Dimension & Meaning \\
  \midrule
  \texttt{NewTopicFit} & Whether the recommendation fits the new interest direction \\
  \texttt{AdaptationAppropriateness} & Whether the system's response to the interest change is appropriate \\
  \texttt{OldNewBalance} & Whether the balance between old and new interests is reasonable \\
  \texttt{DriftDecisionHelpfulness} & Whether the recommendation helps the user decide whether to continue along the new direction \\
  \bottomrule
  \end{tabular*}
  \vspace{-3mm}
\end{table}

Drift human evaluation samples only episodes in \texttt{observing}, \texttt{shifting}, or \texttt{recovered} states, focusing on how the system responds to the new interest direction before and after drift. For finer analysis, one can score recommendations before, during, and after drift separately and compare how human scores change over states.

\subsection{Inter-Annotator Agreement and Correlation Analysis}

We report pairwise Pearson and Spearman correlations among the three annotators.
For the main listwise evaluation, agreement is high across dimensions: mean
pairwise Spearman correlation is 0.923 for ProfileMatch, 0.909 for
RankingQuality, 0.872 for DecisionUsefulness, 0.844 for
DiversityFocusBalance, and 0.901 for OverallRank.

To test alignment between automatic metrics and human judgment, we compute
automatic--human correlations on the sampled blind-evaluation set. For the main
experiment, each point is one sampled method--episode Top-20 list. On the 36
annotated lists, \texttt{RecommendationScore} correlates with
\texttt{HumanEval} at Pearson $r=0.8626$ and Spearman $\rho=0.8631$.
The component metric \texttt{gNDCG@20} also correlates strongly with
RankingQuality, with Pearson $r=0.8723$ and Spearman $\rho=0.8774$.
 
For the drift-specific evaluation, each point is one method--drift-episode
recommendation list. On the 72 annotated lists, \texttt{DriftAutoScore}
correlates with \texttt{AdaptationHumanScore} at Pearson $r=0.9149$ and
Spearman $\rho=0.8904$.

For the model-comparison evaluation, each point is one LLM backbone. Across the
14 completed backbones, \texttt{ModelAutoScore} correlates with
\texttt{ModelHumanScore} at Pearson $r=0.9632$ and Spearman $\rho=0.9648$.
The model-comparison plot uses \texttt{ModelAutoScore} and
\texttt{ModelHumanScore}, while the drift-specific plot uses
\texttt{DriftAutoScore} and \texttt{AdaptationHumanScore}. These analyses check
whether automatic metrics show the same trend as human judgment; they do not
replace the individual metrics in the main result tables.

\FloatBarrier
\section{Real-User Pilot Study Details}
\label{app:real-user-pilot}

\subsection{Study Setup}

The real-user pilot study is designed to complement the simulated benchmark
with a small amount of direct user-experience evidence. Five graduate students
in computer science and AI participate in the study. Each participant uses
PaperFlow for 5--7 interaction rounds. In each round, the system presents a
Top-20 daily paper list, and the participant marks papers they would read,
papers they open or inspect in detail, and optional reading-report requests.

The study is not intended as a statistically powered deployment evaluation.
Instead, it provides a sanity check on whether the ranking and reading-assist
workflow produces useful recommendations for real researchers. All participant
identifiers are anonymized.

\subsection{Participants}

\begin{table}[h]
  \centering
  \caption{Participant information for the real-user pilot study.}
  \label{tab:real-user-participants}
  \small
  \begin{tabular}{@{}c p{0.36\columnwidth} p{0.26\columnwidth} c@{}}
  \toprule
  User & Research Area & Experience & Rounds \\
  \midrule
  U1 & NLP / Summarization & MS-2nd year & 6 \\
  U2 & CV / Video Generation & MS-1st year & 5 \\
  U3 & RecSys / LLM-based Systems & MS-2nd year & 7 \\
  U4 & Multimodal Learning & MS-1st year & 5 \\
  U5 & RL / Code Generation & MS-2nd year & 6 \\
  \bottomrule
  \end{tabular}
\end{table}

\subsection{Metrics}

The real-user pilot uses two behavioral metrics and two questionnaire metrics.
The behavioral metrics are computed from participants' actions on each Top-20
list, while the questionnaire metrics are collected after each interaction round.

For participant $u$ and interaction round $t$, let $R_{u,t}^{(k)}$ denote the
top-$k$ recommended papers. Let $S_{u,t}$ denote the set of papers that the
participant marks as worth reading. This is a broad positive-feedback set: it
captures papers that the participant considers relevant or potentially useful
after seeing the recommendation list. We compute:
\begin{equation}
\mathrm{Prec@}k(u,t)=
\frac{|R_{u,t}^{(k)} \cap S_{u,t}|}{k}.
\end{equation}
Thus, \texttt{Prec@5} measures the concentration of worth-reading papers near
the top of the list, while \texttt{Prec@20} measures the overall usefulness of
the full displayed list.

ReadRate measures a stricter form of engagement. Let $D_{u,t}$ denote the set
of papers that receive an explicit reading action, including opening the paper,
inspecting it in detail, or requesting a reading report. These actions indicate
stronger engagement than simply marking a paper as worth reading. We compute:
\begin{equation}
\mathrm{ReadRate}(u,t)=
\frac{|R_{u,t}^{(20)} \cap D_{u,t}|}{20}.
\end{equation}
Compared with \texttt{Prec@20}, \texttt{ReadRate} is therefore a more
action-oriented metric: \texttt{Prec@20} asks whether the Top-20 list contains
papers the participant would consider reading, whereas \texttt{ReadRate} asks
how many papers actually trigger a concrete reading action.

After each round, participants answer a 1--5 Likert questionnaire.
\texttt{Sat.} is the overall satisfaction score:
\begin{equation}
\mathrm{Sat.}(u,t)=Q_{\mathrm{sat}}(u,t).
\end{equation}
\texttt{ReportUse.} is the reading-report usefulness score and is computed only
for PaperFlow, because the two pilot baselines do not provide reading reports.

For each participant, metric values are first averaged over that participant's
interaction rounds. The final pilot-study results are then averaged over the
five participants. In this sense, \texttt{Prec@5}, \texttt{Prec@20}, and
\texttt{ReadRate} summarize observed behavior, while \texttt{Sat.} and
\texttt{ReportUse.} summarize subjective user experience.

\subsection{Real-User Baselines}

We compare PaperFlow with two lightweight baselines in the pilot study. Daily
arXiv Email corresponds to a date-based paper feed without personalization
beyond broad topic subscription. Static Profile ranks papers using the initial
user profile but does not apply profile updates, reading signals, or drift
adaptation. These baselines are used only for the real-user pilot and are not
intended to replace the controlled benchmark baselines in the main experiment.

\subsection{Questionnaire}

After each interaction round, participants rated the following items on a
1--5 Likert scale.

\begin{table}[h]
  \centering
  \caption{Questionnaire items for the real-user pilot study.}
  \label{tab:real-user-questionnaire}
  \small
  \begin{tabular}{@{}lp{0.74\columnwidth}@{}}
  \toprule
  ID & Question \\
  \midrule
  Q1 & How relevant are today's recommended papers to your current research? \\
  Q2 & How accurately does the system represent your research interests? \\
  Q3 & How useful are the reading reports for understanding selected papers? \\
  Q4 & Compared with previous rounds, did today's recommendations improve? \\
  Q5 & Would you be willing to continue using this system? \\
  Q6 & Overall, how satisfied are you with today's recommendation list? \\
  \bottomrule
  \end{tabular}
\end{table}

\subsection{Per-User Results}

\begin{table}[H]
  \centering
  \caption{Per-user averages in the real-user pilot study.}
  \label{tab:real-user-per-user}
  \small
  \setlength{\tabcolsep}{2.5pt}
  \begin{tabular}{@{}lrrrrr@{}}
  \toprule
  User & Prec@5 & Prec@20 & ReadRate & Sat. & ReportUse. \\
  \midrule
  U1 & 0.72 & 0.48 & 0.35 & 4.1 & 4.3 \\
  U2 & 0.64 & 0.42 & 0.30 & 3.8 & 4.1 \\
  U3 & 0.80 & 0.53 & 0.40 & 4.4 & 4.6 \\
  U4 & 0.68 & 0.45 & 0.33 & 3.9 & 4.0 \\
  U5 & 0.70 & 0.46 & 0.32 & 4.0 & 4.4 \\
  \midrule
  Mean & 0.71 & 0.47 & 0.34 & 4.0 & 4.3 \\
  \bottomrule
  \end{tabular}
\end{table}
\FloatBarrier

\subsection{Ethics and Anonymization}

The pilot study uses anonymized participant identifiers and reports only
aggregate or per-user averaged statistics. Participants are informed that the
study is used to evaluate research-paper recommendation experience, and no
private paper-reading logs are released. The pilot is limited in scale, so we
avoid statistical significance claims and use the results only as complementary
evidence.
\section{Case Studies}
\label{app:case-studies}

\subsection{Case-Study Selection Criteria}

\begin{figure*}[p]
\centering
\makebox[\textwidth][c]{%
\includegraphics[width=1\textwidth,trim=14 8 40 8,clip]{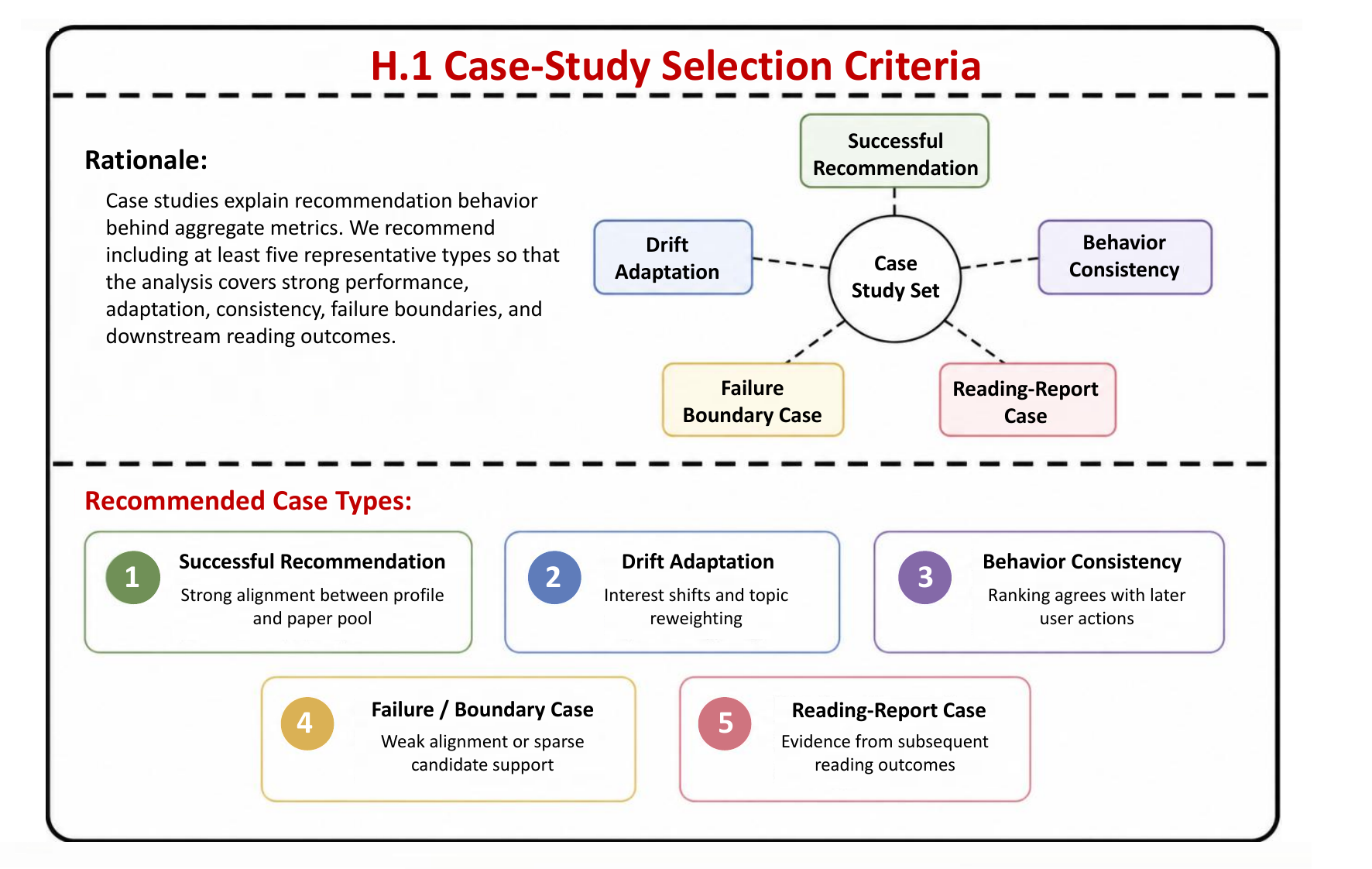}
}
\caption{Case-study selection criteria. The figure summarizes five representative case types for inspecting PaperFlow beyond aggregate metrics: successful recommendation, drift adaptation, behavior consistency, boundary disagreement, and reading-report support.}
\label{fig:case-study-selection-criteria}
\end{figure*}

Case studies complement aggregate metrics with qualitative evidence about why a recommendation list succeeds, adapts, or fails. As summarized in Figure~\ref{fig:case-study-selection-criteria}, we organize the analysis around representative cases covering successful recommendation, interest drift, behavior consistency, boundary conditions, and downstream reading support.

This design avoids selecting only strong outputs. Instead, the case-study set checks whether PaperFlow is useful under different recommendation states, including aligned profile--paper matches, changing user interests, behavior-aware ranking effects, and cases where automatic oracle labels do not fully capture user behavior.

Operationally, these cases are selected from episodes with high ranking quality, drift-state transitions, high behavioral agreement, boundary disagreement between system and oracle labels, or complete PDF-based reading reports.

\FloatBarrier

\subsection{Successful Recommendation: A Dense Top-20 for an NLP User}

\begin{figure*}[p]
\centering
\makebox[\textwidth][c]{%
\includegraphics[width=1.04\textwidth,trim=20 8 20 8,clip]{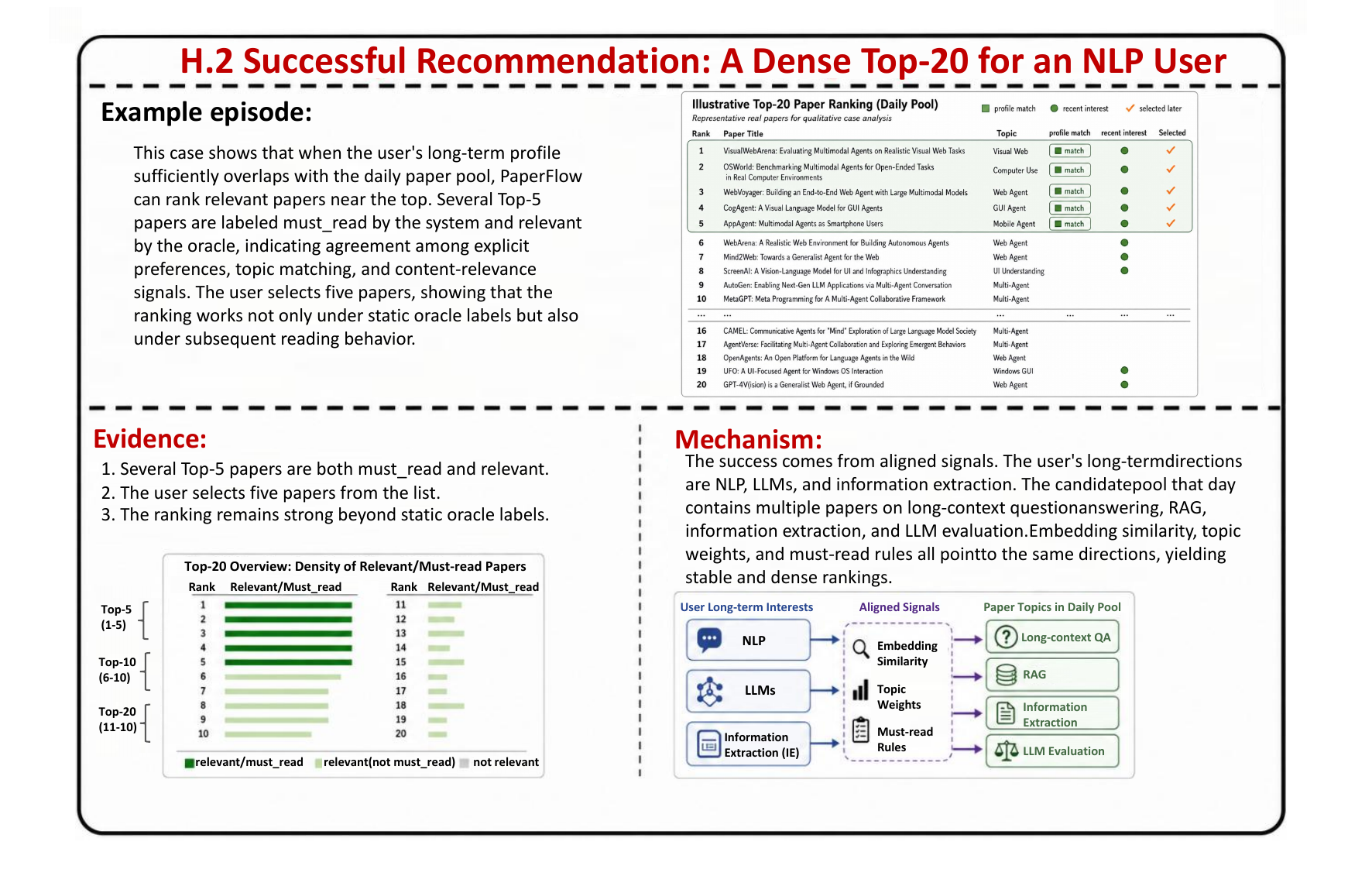}
}
\caption{Successful recommendation case for an NLP user. The episode shows that the user's NLP/LLM/information-extraction profile aligns with the daily candidate pool, producing a dense Top-20 list with useful papers near the top.}
\label{fig:case-nlp-success-visual}
\end{figure*}

This case examines \texttt{user\_role6::2026-03-16}, an NLP user whose directions include NLP, large language models, and information extraction. The episode reaches \texttt{gNDCG@20 = 1.0000}, with 20 useful papers in the Top-20, 10 strictly relevant papers, and five selected papers. Figure~\ref{fig:case-nlp-success-visual} visualizes the Top-20 ranking and the evidence behind the dense high-rank recommendations.

The purpose of the case is to show the mechanism behind a strong list rather than repeat every ranked item in text. Representative top-ranked papers include work on long-context question answering, retrieval-augmented generation, and LLM-supported scientific discovery. When profile match, topic similarity, and must-read rules agree, PaperFlow concentrates useful papers near the top and remains consistent with the user's later selections.

\FloatBarrier

\subsection{Interest Drift: From GUI/Web Agents to Multimodal Reasoning}

\begin{figure*}[p]
\centering
\makebox[\textwidth][c]{%
\includegraphics[width=1.01\textwidth,trim=0 18 18 8,clip]{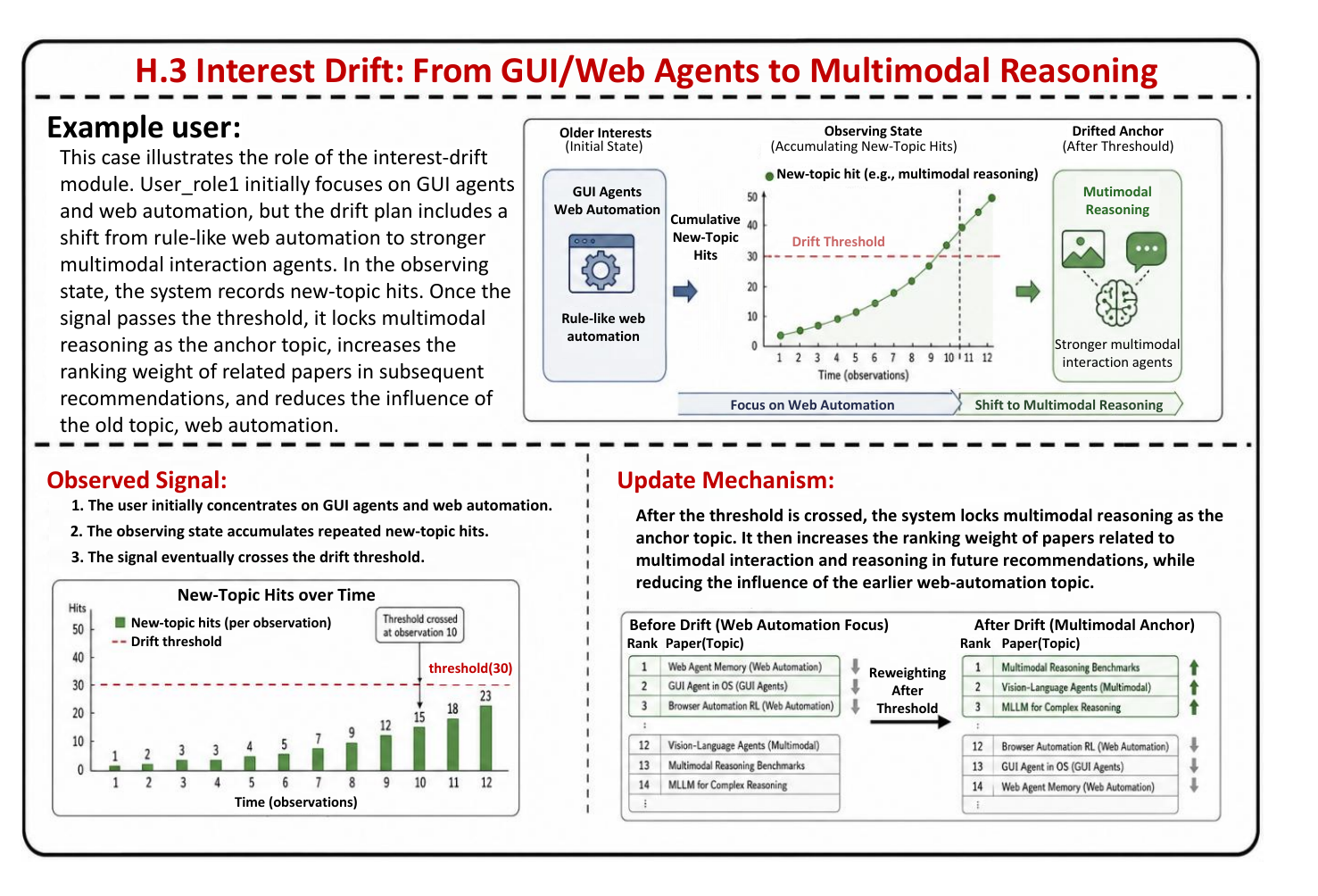}
}
\caption{Interest-drift case from GUI/Web agents to multimodal reasoning. Repeated evidence for a new direction leads PaperFlow to lock a multimodal-reasoning anchor and reweight later recommendations away from stale web-automation interests.}
\label{fig:case-drift-visual}
\end{figure*}

This case focuses on adaptation rather than static relevance. The example user is \texttt{user\_role1}, whose original directions include GUI agents, computer-vision grounding, and web automation. The recorded drift event is an \texttt{anchor\_lock} on 2026-03-02, moving from \texttt{observing} to \texttt{shifting} toward multimodal reasoning and computer-using agents. Figure~\ref{fig:case-drift-visual} summarizes how accumulated new-topic evidence changes the ranking emphasis.

The key observation is that PaperFlow does not immediately replace the user's profile after a single new-topic hit. It waits for repeated evidence, locks the emerging topic as an anchor after the threshold is reached, and then increases exposure to papers aligned with the new direction.

\FloatBarrier

\subsection{Behavior Consistency: A High-\texttt{SelectedNDCG} List}

\begin{figure*}[p]
\centering
\makebox[\textwidth][c]{%
\includegraphics[width=1.01\textwidth,trim=0 8 18 8,clip]{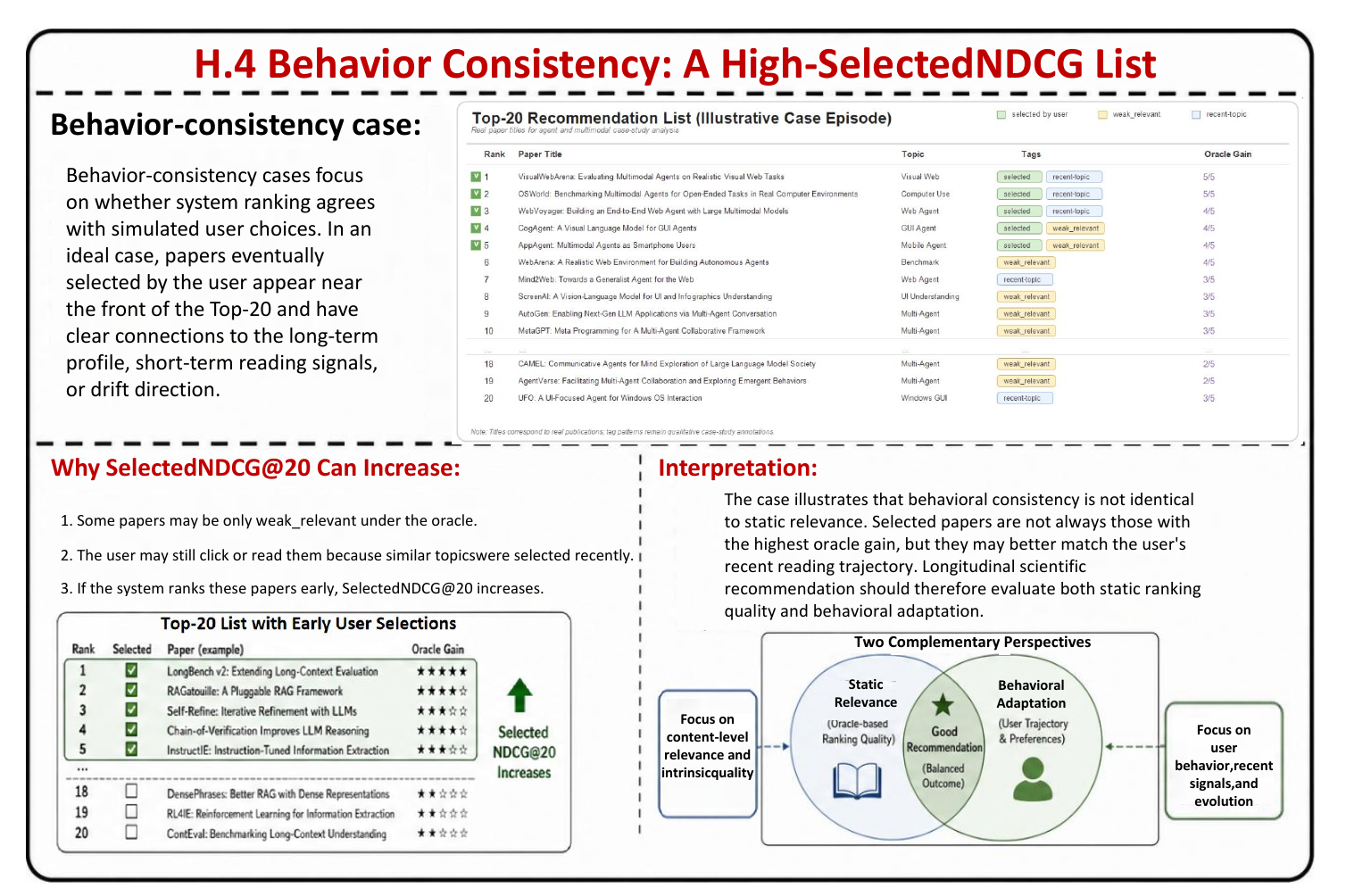}
}
\caption{Behavior-consistency case with a high-\texttt{SelectedNDCG} list. Selected papers appear near the front of the Top-20, showing that behavior-based agreement can complement static oracle relevance in longitudinal recommendation evaluation.}
\label{fig:case-behavior-consistency-visual}
\end{figure*}

Behavior-consistency cases focus on whether system ranking agrees with simulated user choices. Figure~\ref{fig:case-behavior-consistency-visual} shows a high-\texttt{SelectedNDCG} list in which papers selected by the user appear near the front of the Top-20 and have clear connections to the long-term profile, short-term reading signals, or drift direction.

This case follows the behavior-consistency criterion: selected papers should appear near the top or match emerging interests. It clarifies the difference between static relevance and behavioral adaptation. \texttt{SelectedNDCG@20} rewards agreement with later user choices, while oracle labels measure content-level relevance. Qualitative inspection is therefore useful for interpreting cases where a paper is only weakly relevant under the oracle but remains consistent with the user's recent reading trajectory.

\FloatBarrier

\subsection{Failure or Boundary Case: Disagreement Between System Labels and Oracle Labels}

Example episode:

\begin{table}[t]
  \small
  \centering
  \vspace{-1mm}
  \caption{Boundary case with disagreement between system and oracle labels.}
  \vspace{-1mm}
  \label{tab:case-boundary}
  \setlength{\tabcolsep}{1.5mm}
  \begin{tabular*}{\columnwidth}{@{\extracolsep{\fill}}p{0.30\columnwidth}p{0.60\columnwidth}@{}}
  \toprule
  Field & Content \\
  \midrule
  Episode & \texttt{user\_role3::2026-03-24} \\
  User directions & literature mining, scientific knowledge graph, hypothesis generation \\
  Daily result & \texttt{gNDCG@20 = 0.0000}; 0 useful papers in Top-20 \\
  User selections & Number of selected papers = 6 \\
  Top-ranked examples & \textit{Permutation-Symmetrized Diffusion for Unconditional Molecular Generation}; \textit{Describe-Then-Act}; \textit{Off-Policy Value-Based Reinforcement Learning for Large Language Models} \\
  \bottomrule
  \end{tabular*}
  \vspace{-3mm}
\end{table}

This is a typical boundary case. The system may rank some papers highly because of short-term behavior signals, explicit rules, or topic-word matches, while the oracle still labels them as \texttt{irrelevant}. At the same time, the simulated user selects several papers, showing that selection behavior and oracle relevance are not identical. The paper should not describe such a case as a complete system failure. Instead, it reveals possible discrepancies among automatic oracle labels, behavior-selection signals, and system labels, which is one reason human evaluation is needed.

The case also suggests that relevance in scientific recommendation is layered. Some papers may not directly belong to a user's core direction, but may still be inspiring because of their method, data structure, or task formulation. Automatic oracle labels may not fully capture such cross-domain inspiration. Human evaluation and case studies can therefore complement automatic metrics.

\FloatBarrier

\subsection{Reading Report: PDF-Evidence-Driven Reading Assistance}

Example report:

\begin{table}[t]
  \small
  \centering
  \vspace{-1mm}
  \caption{PDF-evidence-driven reading-report case.}
  \vspace{-1mm}
  \label{tab:case-reading-report}
  \setlength{\tabcolsep}{1.5mm}
  \begin{tabular*}{\columnwidth}{@{\extracolsep{\fill}}p{0.30\columnwidth}p{0.60\columnwidth}@{}}
  \toprule
  Field & Content \\
  \midrule
  User & \texttt{user\_role1} \\
  Date & 2026-03-01 \\
  Paper & \textit{The Informational Cost of Agency: A Bounded Measure of Interaction Efficiency for Deployed Reinforcement Learning} \\
  arXiv ID & \texttt{2603.01283v2} \\
  Report source & PDF \\
  Report fields & one-sentence summary, research background, core method, key results, contributions, limitations, relation to user research, reading suggestions, and PDF evidence anchors \\
  \bottomrule
  \end{tabular*}
  \vspace{-3mm}
\end{table}

This case shows that PaperFlow does not only output a recommendation list. It also converts selected papers into structured reading-assistance material. Report generation combines the abstract, PDF sections, semantic-retrieval evidence, and user profile, so it can explain why the paper is relevant to the user, which parts should be read first, and how the paper might support the user's current research. The main paper can show only a short excerpt and place the full report in the appendix.

From the system-value perspective, reading reports distinguish PaperFlow from ordinary paper recommenders. Traditional systems usually provide only titles, abstracts, and relevance scores. PaperFlow additionally generates profile-aware reading guidance. For researchers, this can reduce the reading cost after paper screening.

\section{Model Comparison Details}
\label{app:model-comparison-details}

\subsection{Purpose of Model Comparison}

The LLM comparison evaluates how different large language models behave inside the PaperFlow framework. Since the recommendation pipeline includes LLM-supported profile parsing, recommendation-explanation generation, and reading-report generation, model differences in structured judgment, long-context understanding, reading-report generation, and cost can affect overall system performance.

The model-comparison experiment does not construct a new benchmark. It replaces the language model within the same benchmark snapshot. This ensures that different models face the same users, the same candidate pools, and the same evaluation metrics. Results report \texttt{Recommendation\allowbreak Score},
\texttt{Report\allowbreak Auto\allowbreak Score},
\texttt{Model\allowbreak Auto\allowbreak Score}, and
\texttt{Model\allowbreak Human\allowbreak Score}; \texttt{Token\allowbreak Cost}
is reported separately as an efficiency metric and is not included in the
quality score.

\texttt{ReportAutoScore} follows the definition in the model-comparison
section and combines SectionCompleteness and EvidenceCoverage. Operationally,
SectionCompleteness corresponds to \texttt{ReportStructureScore}, and
EvidenceCoverage corresponds to \texttt{ReportEvidenceRate}.

\texttt{ParsingSuccess} measures structured-output stability, but all completed
model runs achieve 100\% non-empty report generation success.
It is therefore omitted from the main model-comparison table.

The cost-free \texttt{ModelAutoScore} follows the definition in the
model-comparison section and is used as the quality score in the model-comparison
table. \texttt{TokenCost} is derived from token-usage logs and reported
separately as an efficiency metric.

\FloatBarrier

\subsection{Open and Closed Model Scope}

The model-comparison study uses only the completed model runs reported in
Table~\ref{tab:llm-comparison}; candidate models that were planned but not run
are excluded. We group the backbones by access mode. The closed API group
contains GPT-5.4, Qwen3.5-Plus, Gemini 3.1 Pro Preview, Claude Sonnet 4.6,
Qwen3.6-Plus, Qwen3.6-Max-Preview, Grok 4.3, and the default PaperFlow setting
based on Gemini 3 Flash Preview. The open/open-access group contains
MiMo-V2.5-Pro, DeepSeek-V4-Pro, DeepSeek-V4-Flash, Kimi K2.6, GLM-5.1, and
MiniMax-M2.7.

All models are evaluated with the same benchmark snapshot, embedding model,
Top-20 display budget, and evaluation metrics.\texttt{TokenCost} is
reported only as an efficiency metric and is not included in
\texttt{ModelAutoScore} or \texttt{ModelHumanScore}.

\subsection{Model-Comparison Controls}

The model comparison controls the following variables:

\begin{enumerate}
\item The candidate paper pool remains fixed.
\item User profiles remain fixed.
\item The Top-20 display budget remains fixed.
\item The embedding model remains fixed.
\item Evaluation metrics remain fixed.
\item Each model writes to a separate output directory.
\item Token usage is recorded by date.
\item \texttt{TokenCost} is reported only as an efficiency metric and is not included in \texttt{ModelAutoScore} or \texttt{ModelHumanScore}.
\end{enumerate}

If a model encounters JSON parsing errors, API connection errors, or token-usage accounting anomalies, these should be recorded separately and should not be directly compared with models that completed full runs. Missing results caused by model-call failures should be marked as incomplete or unstable in the paper.

\subsection{Token Usage Records}

Each model experiment records daily token usage, including embedding tokens, LLM tokens, total tokens, and call count. Variation in embedding tokens is usually related to candidate-pool embedding cache status. If embedding tokens are zero on some dates, the corresponding embeddings may already be cached. LLM tokens mainly come from structured parsing, recommendation explanations, and reading-report generation.

Token usage can support cost analysis and explain runtime differences across models. A full benchmark contains 1200 episodes and reading reports for selected papers, so total runtime and token usage can be substantial.

\section{Reproducibility Notes}
\label{app:reproducibility-notes}

\subsection{Runtime Environment}

The experiment runs in a Python environment and uses a SQLite database to store user profiles, the paper table, behavior logs, and task state. Each model package contains independent run scripts, configuration files, result directories, and token-usage logs.

Before a full run, the system resets benchmark state while retaining the papers table. This removes user behavior, task state, and result records generated by a previous model run without rebuilding the paper pool.

\subsection{Output Files}

Each full experiment produces multiple types of output files.

\begin{table*}[t]
  \centering
  \caption{Runtime records and output files.}
  \label{tab:runtime-records-output-files}
  \small
  \setlength{\tabcolsep}{4pt}
  \renewcommand{\arraystretch}{1.08}
  \begin{tabular*}{\textwidth}{@{\extracolsep{\fill}}p{0.18\textwidth}p{0.27\textwidth}p{0.47\textwidth}@{}}
  \toprule
  Category & Field & Meaning \\
  \midrule
  Token usage & \texttt{date} & Date \\
  Token usage & \texttt{embedding\_tokens} & Embedding tokens used on that day \\
  Token usage & \texttt{llm\_tokens} & LLM tokens used on that day \\
  Token usage & \texttt{total\_tokens} & Sum of embedding and LLM tokens \\
  Token usage & \texttt{call\_count} & Number of API calls that day \\
  \midrule
  Runtime environment & Database & SQLite \\
  Runtime environment & Main script & \texttt{scripts/simulate\_historical\_episodes.py} \\
  Runtime environment & Database cleanup script & \texttt{scripts/clear\_database.py} \\
  Runtime environment & Result directory & \texttt{results/\{model\_key\}} \\
  Runtime environment & Token log & \texttt{results/\{model\_key\}/token\_usage.jsonl} \\
  Runtime environment & Reading-report directory & \texttt{results/\{model\_key\}/reading\_reports\_md} \\
  \midrule
  Output files & Episode metadata & Recommendation context and results for each user-day episode \\
  Output files & Ranking results & Top-20 recommendation lists and scores \\
  Output files & Behavior logs & Simulated user-selection behavior \\
  Output files & Reading reports & Reading reports for selected papers \\
  Output files & Token usage & Daily token usage \\
  Output files & Drift timeline & Interest-drift state-change records \\
  Output files & Aggregate metrics & Aggregated main metrics and model-comparison results \\
  \bottomrule
  \end{tabular*}
  \vspace{-2mm}
\end{table*}

These outputs are used for later statistics, table generation, case-study selection, and human-evaluation sample construction.

\subsection{Error Handling}

External API or network errors can occur during the experiment, including arXiv API timeout, 429 rate limits, PDF download failure, LLM connection error, or JSON parsing error. The system provides fallback mechanisms for some errors. For example, if arXiv metadata retrieval fails, it attempts to download the PDF; if PDF retrieval fails, it uses the title and abstract to generate a simplified report.

Two types of errors should be distinguished. The first type is recoverable, such as an arXiv timeout followed by successful PDF parsing; this usually does not affect the main flow. The second type is unrecoverable, such as path-length write failure, complete LLM API unavailability, or unparseable output; these require code or configuration fixes and rerunning the experiment.

In the paper, if a model experiment is incomplete because of API instability, it should be explicitly marked as incomplete rather than directly comparing partial results with complete runs.

\clearpage

\end{document}